\documentclass[longauth]{aa}                     
\usepackage{graphicx}
\usepackage{amsmath}
\usepackage{xcolor}
\usepackage{url}
\usepackage{hyperref}

\UseRawInputEncoding

\usepackage{txfonts}
\usepackage[normalem]{ulem}

\newcommand{\bhb}{[BHB2007] 11}
\newcommand{\chtoh}{CH$_3$OH}
\newcommand{\chtocho}{CH$_3$OCHO}
\newcommand{\chtocht}{CH$_3$OCH$_3$}
\newcommand{\sot}{SO$_2$}

\begin{document}

\titlerunning{FAUST XI: Enhancement of the complex organic material in shocked matter}
   \title{FAUST XI: Enhancement of the complex organic material in the shocked matter surrounding the \bhb~protobinary system}

  \author{C. Vastel\inst{1}
        \and
        T. Sakai\inst{2}
        \and
        C. Ceccarelli\inst{3}
        \and
        I. Jim\'{e}nez-Serra\inst{4}
        \and
        F. Alves\inst{5}
        \and
        N. Balucani\inst{6}
        \and
        E. Bianchi\inst{7}  
        \and
        M. Bouvier\inst{8}
        \and
        P. Caselli\inst{5}      
        \and
        C. J. Chandler\inst{9}
        \and
        S. Charnley\inst{10}
        \and
        C. Codella\inst{11}
        \and
        M. De Simone\inst{12}  
        \and
        F. Dulieu\inst{13}
        \and
        L. Evans \inst{14}
         \and
        F. Fontani\inst{11}
        \and
        B. Lefloch\inst{15}
        \and
        L. Loinard\inst{16}
        \and
        F. Menard\inst{3}
        \and
        L. Podio\inst{11}
        \and
        G. Sabatini \inst{11}
        \and
        N. Sakai\inst{17}
        \and
        S. Yamamoto\inst{18,19}
 }

   \institute{IRAP, Universit\'{e} de Toulouse, CNRS, CNES, UPS, Toulouse, France\\
              \email{cvastel@irap.omp.eu}
         \and
            Graduate School of Informatics and Engineering, The University of Electro-Communications, Chofu, Tokyo 182-8585, Japan
        \and
            Universit\'{e} Grenoble Alpes, CNRS, Institut de Plan\'{e}tologie et d'Astrophysique de Grenoble (IPAG), 38000 Grenoble, France
        \and
            Centro de Astrobiologia (CSIC-INTA), Ctra. de Torrejon a Ajalvir, km 4, 28850, Torrejon de Ardoz, Spain
         \and
            Max-Planck-Institut f\"{u}r extraterrestrische Physik (MPE), Gie{\ss}enbachstr. 1, D-85741 Garching, Germany
         \and
           Department of Chemistry, Biology, and Biotechnology, The University of Perugia, Via Elce di Sotto 8, 06123 Perugia, Italy
         \and
           Excellence Cluster ORIGINS, Boltzmannstra{\ss}e 2, D-85748 Garching bei M\"{u}nchen, Germany; Ludwig-Maximilians-Universit\"{a}t, Schellingstra{\ss}e 4, D-80799 M\"{u}nchen, Germany
        \and
           Leiden Observatory, Leiden University, P.O. Box 9513, 2300 RA Leiden, The Netherlands
        \and
           National Radio Astronomy Observatory, PO Box O, Socorro, NM 87801, USA
        \and
           Astrochemistry Laboratory, Code 691, NASA Goddard Space Flight Center, 8800 Greenbelt Road, Greenbelt, MD 20771, USA   
        \and
            INAF-Osservatorio Astrofisico di Arcetri, Largo E. Fermi 5, I-50125, Florence, Italy
        \and
            European Southern Observatory, Karl-Schwarzschild Str. 2, 85748 Garching bei M\"{u}nchen, Germany
        \and
           Cergy Paris Universit\'e, Sorbonne Universit\'e, Observatoire de Paris, PSL University, CNRS, LERMA, F-95000, Cergy, France     
        \and
           School of Physics and Astronomy, University of Leeds, Leeds, LS2 9JT, UK 
        \and
           Laboratoire d'Astrophysique de Bordeaux, Univ. Bordeaux, CNRS, B18N, all\'ee Geoffroy Saint-Hilaire, 33615, Pessac, France     
        \and
           Instituto de Radioastronomia y Astrofisica, Universidad Nacional Aut\'onoma de Mexico Apartado, 58090, Morelia, Michoac\'an, Mexico         
        \and
            RIKEN Cluster for Pioneering Research, 2-1, Hirosawa, Wako-shi, Saitama 351-0198, Japan
        \and
           The Graduate University for Advanced Studies SOKENDAI, Shonan Village, Hayama, Kanagawa 240-0193, Japan       
        \and 
            Research Center for the Early Universe, The University of Tokyo, 7-3-1, Hongo, Bunkyo-ku, Tokyo 113-0033, Japan     
}

   \date{Received November 11th 2023; accepted February 8th 2024}

 
  \abstract
{}
{Interstellar complex organic molecules (iCOMs) are species commonly found in the interstellar medium. They are believed to be crucial seed species for the build-up of chemical complexity in star forming regions as well as our own Solar System. Thus, understanding how their abundances evolve during the star formation process and whether it enriches the emerging planetary system is of paramount importance. }
{We use data from the ALMA Large Program FAUST (Fifty AU STudy of the chemistry in the disk and envelope system of solar protostars) to study the compact line emission towards the \bhb~proto-binary system (sources A and B), where a complex structure of filaments connecting the two sources with a larger circumbinary disk has previously been detected. More than 45 methyl formate (\chtocho) lines are clearly detected with upper energies in the [123, 366] K range, as well as 8 dimethyl ether transitions (\chtocht) in the [93, 191] K range, 1 ketene transition (H$_{2}$CCO) and 4 formic acid transitions (t-HCOOH). We compute the abundance ratios with respect to \chtoh~for \chtocho, \chtocht, H$_{2}$CCO, t-HCOOH (as well as an upper limit for CH$_{3}$CHO) through a radiative transfer analysis. We also report the upper limits on the column densities of nitrogen bearing iCOMs, $N$(C$_2$H$_5$CN) and $N$(C$_2$H$_3$CN).}
{The emission from the detected iCOMs and their precursors is compact and encompasses both protostars, which are separated by only 0.2$^{\prime\prime}$ ($\sim$ 28 au). The integrated intensities tend to align with the Southern filament, revealed by the high spatial resolution observations of the dust emission at 1.3 mm. 
A Position-Velocity and 2D analysis are performed on the strongest and uncontaminated \chtocht~transition and show three different spatial and velocity regions, two of them being close to 11B (Southern filament) and the third one near 11A. }
{All our observations suggest that the detected methanol, as well as the other iCOMs, are generated by the shocked gas from the incoming filaments streaming towards [BHB2007] 11A and 11B, respectively, making this source one of the few where chemical enrichment of the gas caused by the streaming material is observed.}
{}

 \keywords{astrochemistry - ISM: abundance - ISM: molecules - line: identification}

 \maketitle


\section{Introduction}
\label{sec: Intro}

Interstellar complex organic molecules (also called iCOMs)  are defined in the astronomical community as carbon-bearing species containing at least six atoms \citep{Herbst2009,Ceccarelli2017}. The detection of iCOMs in solar-type protostars and their precursors in the prestellar core phase is crucial for understanding how molecular complexity emerges and evolves in the different stages that lead to the formation of a solar system such our own. The protostellar phase starts with the young accreting protostar phase with a central object  at T $\ge$ 100 K that eventually will become a star,  surrounded by a colder envelope from which the future star accretes matter (also called Class 0). In a simple scenario, the molecules trapped in the ices are injected back into the gas phase due to thermal desorption, giving rise to a rich and peculiar chemistry in which iCOMS are found in these regions, called hot corinos \citep{Cazaux2003,Bottinelli2004,Ceccarelli2004}. The name comes from their high-mass analogues, which are called hot cores \citep{Morris1980}. Many iCOMS have been detected in these objects, such as methyl formate (\chtocho), dimethyl ether (\chtocht), acetaldehyde (CH$_3$CHO), ethanol (CH$_3$CH$_2$OH), and formamide (NH$_2$CHO) \citep[see, e.g., the recent review by][]{Ceccarelli2023}. \\
In this context, the Atacama Large Millimeter/submillimeter Array (ALMA) Large Program\footnote{\url{http://faust-alma.riken.jp}} Fifty AU STudy of the chemistry in the disk and envelope system of Solar-like protostars: 2018.1.01205.L (FAUST) is designed to survey the chemical composition of a sample of 13 Class 0/I protostars at a scale of about 50 to 1000 au (all sources have a distance smaller than 250 pc). The description of the project is reported in \citet{Codella2021}. 
One of these sources is the \bhb~protobinary system, which is embedded in the darkest parts of the Pipe nebula. This is a quiescent complex of interstellar clouds located at a distance of 163 $\pm$ 5 pc \citep{Dzib2018} from the Sun. Only a handful of embedded sources are known in the entire complex in the B59 molecular cloud, with \bhb~being the youngest member of a protocluster of low-mass young stellar objects \citep{Brooke2007}. At a higher angular resolution ($\sim$ 0.04$^{\prime\prime}$), \citet{Alves2019} previously uncovered a young binary protostar system embedded in circumstellar disks with radii of 2 to 3 au. The protostars are named \bhb~A (hereafter 11A) for the northern source and \bhb~B (hereafter 11B) for the southern source. Both cores are surrounded by a complex filamentary structure (within $\sim$ 100$^{\prime\prime}$) connecting to the larger circumbinary disk. \citet{Alves2019} suggested that these filaments are inflow streamers (also called accretion streamers) from the extended circumbinary disk onto the circumstellar disks of the protobinary system. Their CO observations primarily trace the outflow emission, but the high-velocity components indicate acceleration of infalling gas from the circumbinary disk onto \bhb~B. No high-velocity components are detected near \bhb~A. \citet{Evans2023} recently presented large-scale formaldehyde emission that seems to be consistent with an asymmetric molecular outflow launched by a wide-angle disk wind. 
\citet{Vastel2022} reported the emission of a compact hot methanol emission with three velocity components at -2, 2.8, and 9.9 km~s$^{-1}$ and surprisingly found that the methanol components are not associated with the dust emission peaks, which would suggest a hot-corino origin. An estimation of the dust opacities using the 225 GHz dust continuum emission, which could lead to a high attenuation of the methanol emission at the dust peak, had uncertainties that were too large for any conclusive evidence of whether the foreground dust could completely attenuate the methanol emission from the hot corinos. Another possibility would be that methanol emission originates from the shocked positions at the intersection of intertwined filaments, leading to a different origin than the classical hot-corino origin. This could result in the sputtering of the grain mantles in these shocked regions due to the streaming of the gas towards 11A and 11B, which impacts the quiescent gas of the circumbinary envelope or, possibly, the two circumstellar disks \citep{Vastel2022}. \citet{Evans2023} also presented some HDCO and D$_2$CO emission that was only found at the same location of the methanol compact emission, deriving an average D/H ratio of 0.02$^{+0.02}_{-0.01}$.\\
In the present work, we investigate the iCOM emission found in the FAUST observations towards \bhb~as well as some precursors. Our goal is to understand the origin of the detected iCOM emission by locating and comparing it with the most simple iCOM, methanol, which was previously studied by \citep{Vastel2022}. We present in Section \ref{sec: obs} the observations and describe in Section \ref{sec: line-id} the method we used for the line identification of the iCOMs and precursors. Section \ref{sec: results} presents the spatial distribution of isolated transitions in order to determine the location of their emission. To compute the column densities of the species of interest, we perform a radiative transfer analysis of the emission lines in Section \ref{sec: modelling}. A discussion is finally developed in Section \ref{sec: discussion} to understand the origin of the emission of the iCOMs in \bhb~and compare with other protostellar sources.

\section{Observations}
\label{sec: obs}

\begin{figure*}
    \centering
    \includegraphics[width = \textwidth,trim=0 0 0 0 clip]{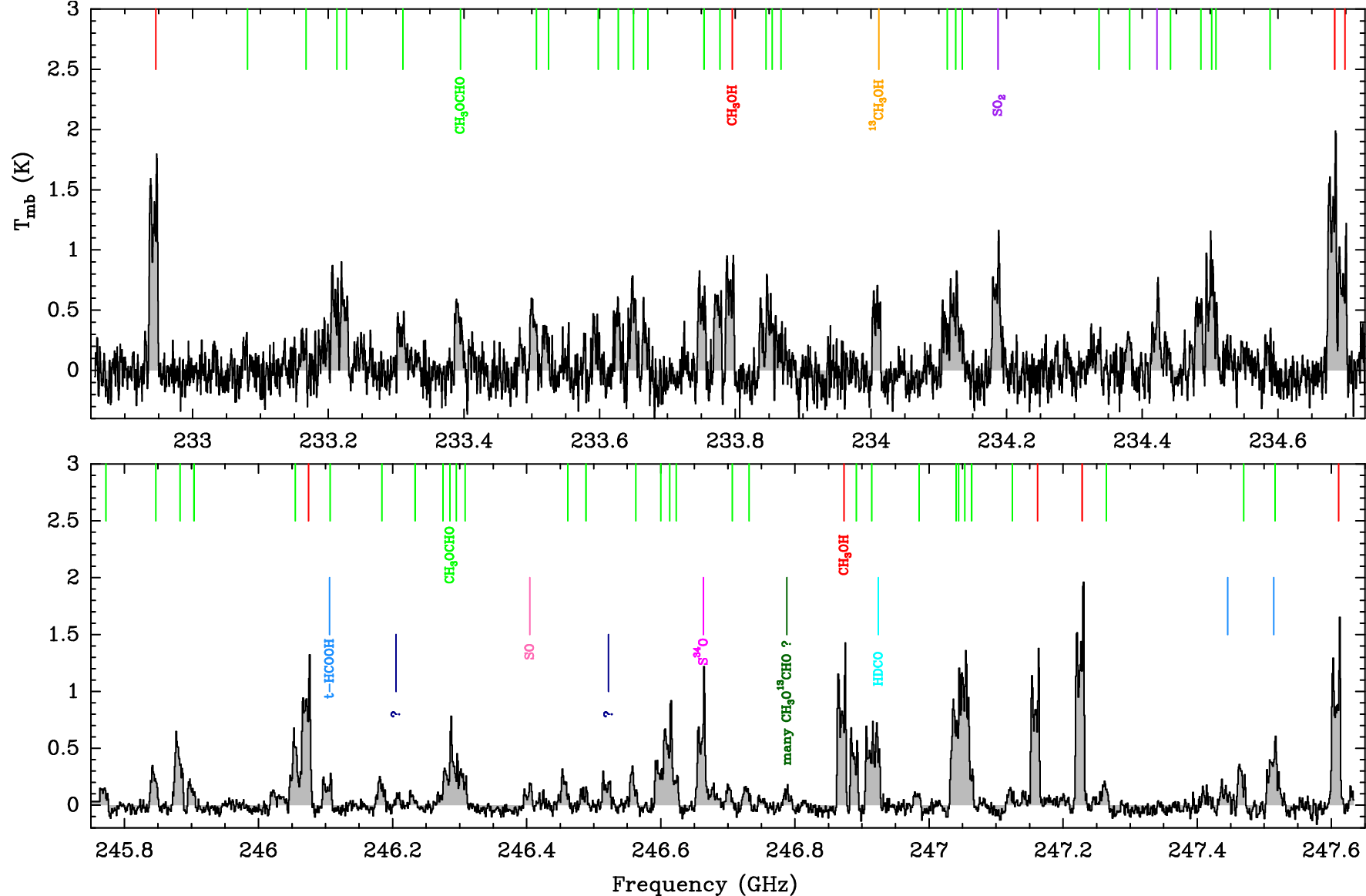}
    \caption{Spectra extracted within $0.82^{\prime\prime} \times 0.77^{\prime\prime}$ from the two continuum bands (see Table \ref{tab: spw}). The grey shaded area corresponds to the filled histogram above and below the baseline. All the green lines correspond to the methyl formate transitions. Methanol is indicated in red and formic acid in blue.}
    \label{fig: cont-spectra}
\end{figure*}

The observations were centred at $\alpha$ (2000) = 17h11m23.125s, $\delta$ (2000) = -27$^{\circ}$24$^{\prime}$32.87$^{\prime\prime}$. They consisted of three frequency setups and 13 spectral windows ({\it spw}), as explained in previous FAUST papers on \bhb~\citep{Vastel2022,Evans2023}. Table \ref{tab: spw} presents the {\it spw} for the FAUST \bhb~observations, specifically, those in which detections are presented in this study (in bold), as well as those used to compute the upper limits. All the $spw$ were used to search for transitions of the species of interest. We used the 12m array (TM 1 and TM 2) due to the compact emission of the iCOMs transitions.  The bandwidth is 59 MHz for the 12m array, except for the continuum window, with a 1.875 GHz bandwidth. Spectral line imaging was performed with the Common Astronomy Software Applications package (CASA) 5.6.1-8 \footnote{\url{https://casa.nrao.edu/}}  with a modified version of the ALMA calibration pipeline and additional in-house calibration routine\footnote{\url{https://faust-imaging.readthedocs.io/en/latest/}} (Chandler in prep.). The systemic velocity of \bhb~is 3.6 km~s$^{-1}$ with respect to the local standard of rest \citep{Onishi1999}. For the following line identification, we extracted two spectra from the {\it spw} continuum bands to cover the broad velocity emission, which extends within 0.82$^{\prime\prime}$ $\times$ 0.77$^{\prime\prime}$ centred at $\alpha$ (2000) = 17h11m23.107s, $\delta$ (2000) = -27$^{\circ}$24$^{\prime}$33.005$^{\prime\prime}$. The two spectra are presented in Fig~\ref{fig: cont-spectra}, and the resulting beams and spectral resolutions are quoted in Table \ref{tab: spectro}.
Based on the high spatial resolution observations of the dust emission at 1.3 mm with ALMA by \cite{Alves2019}, the coordinates for the two cores are $\alpha$ (2000) = 17h11m23.1058s, $\delta$ (2000) = -27$^{\circ}$24$^{\prime}$32.828$^{\prime\prime}$ for 11A  and $\alpha$ (2000) = 17h11m23.1015s, $\delta$ (2000) =  -27$^{\circ}$24$^{\prime}$32.987$^{\prime\prime}$ for 11B. There was a small error in the coordinates, as quoted in the Introduction of \citet{Vastel2022}.

\begin{table}
\small
\centering
\caption{Spectral windows presented in this study for the detected iCOMs (in bold) in the observed FAUST \bhb~dataset. }
\label{tab: spw}
\begin{tabular}{cccc}
\hline
Spectral window   & setup 1                         &  setup 2             & setup 3\\
                             &  (GHz)                              &  (GHz)                   &  (GHz)    \\   
\hline
spw 25   & 216.113                  &  243.916     & 93.181 \\
spw 27   & 216.563                  &  244.049     & 94.405\\
spw 29   & 217.105                  &  244.936     & 95.000$^a$ \\
spw 31   & 217.822                  &  245.606     &  107.014\\
spw 33   & 218.222                  &  {\bf 246.700}$^a$     &  108.040\\
spw 35   & 218.440                  &  {\bf 260.189}     & 104.239 \\
spw 37   & 219.560                  &  260.255     &  105.799\\
spw 39   & 219.949                  & 261.687     & \\
spw 41   & 231.061                  & 262.004     & \\
spw 43   & 213.221                  & {\bf 257.896}     &  \\
spw 45   & 231.322                  & 258.256      & \\
spw 47   & 231.410                  & {\bf 258.549}     &  \\
spw 49  &  {\bf 233.796}$^a$  & {\bf 259.035} \\
\hline
\end{tabular}
\tablefoot{$^a$: continuum windows. The {\it spw} that are not indicated in bold were also used in the search for iCOMs transitions.}
\end{table}

\section{Line identification}
\label{sec: line-id}

In this section, we describe the line identification method for the exploration of the molecular complexity in \bhb. The line identification was performed using the CASSIS\footnote{\url{http://cassis.irap.omp.eu}} software, which connects to the JPL \citep{Pickett1998} and CDMS \citep{Muller2005} databases with direct access to VAMDC\footnote{\url{http://www.vamdc.org/}} as well to verify all the available spectroscopic databases.

\subsection{Methyl formate (\chtocho)}
The large bandwidth of the continuum window led to the detection of more than 45 transitions of methyl formate in setup 1 at 233.796 GHz and setup 2 at 259.035 GHz. We also tentatively identified four CH$_{3}$O$^{13}$CHO transitions blended in one broad feature (see discussion in Sec. \ref{sec: modelling}). Table \ref{tab: spectro} lists the spectroscopic parameters for the lines found in the two continuum bands as well as the transitions detected in the other {\it spw}. In the interstellar medium, methyl formate is the most abundant of its isomers, namely acetic acid (CH$_{3}$COOH, TAG=60523) and glycolaldehyde (HCOCH$_{2}$OH, TAG=60006), which have not been detected in our observations.

\subsection{Dimethyl ether (\chtocht)}
Two strong lines of dimethyl ether were detected at 257.9 GHz and 258.5 GHz (see Table \ref{tab: spectro}). No lines were detected in the continuum {\it spw}, probably because the Einstein coefficients are too low to detect any line above the noise: $A_{ul}$ $\le$ 10$^{-5}$ s$^{-1}$ at 233 GHz and $A_{ul}$ $\le$ 10$^{-8}$ s$^{-1}$ at 246 GHz. In addition, the detected 258.5 GHz line is affected by the emission of a blended transition of \chtocho~(listed in Table \ref{tab: spectro}). 

\subsection{Formamide (NH$_{2}$CHO) and ketene (H$_{2}$CCO)}
Formamide (NH$_{2}$CHO) was targeted in {\it spw} 35 of setup 2 through its 12$_{2,10}$--11$_{2,9}$ transition at 260.190 GHz (see Table \ref{tab: spectroND}). A line is indeed present at the quoted frequency at a 0.1 K level. To confirm the detection, we checked for additional transitions with similar spectroscopic parameters (Einstein coefficient, upper energy level, and statistical weight). One transition at 247.391 GHz should present a similar peak intensity given the similar Einstein coefficient, upper level energy, and degeneracy, but it has not been detected at a 4$\sigma$ level (0.07 K). Similarly, two other transitions at 234.316 GHz and 233.897 GHz (see Table \ref{tab: spectroND} for the spectroscopic parameters) are not detected either at a 0.1 K $rms$. 
On the other hand, ketene (H$_{2}$CCO) seems the most likely candidate with its transition at 260.192 GHz (see Table \ref{tab: spectro}). Unfortunately, there are no other potential transitions for ketene in the continuum {\it spw}, and it is the only transition that is covered within our observations. The line width and $V_{LSR}$ are compatible with \chtoh, \chtocht,~and \chtocho~lines at -2, 2.8, and 9.9 km~s$^{-1}$, and we can safely conclude that the species is detected in \bhb. Although it is not an iCOM in the strict sense, it is thought to be involved in grain-surface reactions to form iCOMs such as formic acid, ethanol, and acetaldehyde \citep[][and references therein]{Charnley1997,Charnley2008,Garrod2008,Hudson2013,Ferrero2023}.

\subsection{Formic acid (t-HCOOH)}
We also detected formic acid in its trans form. However, only four detectable lines lie in our frequency coverage, three of which are blended with CH$_3$OCHO at 246.1 and 247.5 GHz. Although it is not an iCOM by the above strict definition, t-HCOOH is important as it contains three heavy atoms and
is the simplest organic acid. The spectroscopic parameters are listed in Table \ref{tab: spectro}.

\subsection{Unidentified lines}
In addition, two unidentified transitions have been detected at 246.522 GHz and 246.209 GHz that do not correspond to any firm detected transition, according to the CDMS and JPL databases. For the first transition, a CH$_{3}$CDO (TAG=45524) transition (13$_{3,11}$ -- 12$_{3,10}$, $E_{up}$ = 98.14 K, $A_{ul}$ = 5.01~10$^{-4}$ s$^{-1}$) lies at that exact frequency (246522.1360 MHz). However, no other transitions with similar $E_{up}$ and $A_{ul}$ were detected in the continuum {\it spw} within the noise. For the second transition, a c-$^{13}$CCCH (TAG=38005) transition (N$_{K_{-1},K_{+1}}$ = 5$_{1,4}$--4$_{1,3}$, J = 9/2--7/2, F1 = 5--4, F = 11/2--9/2, $E_{up}$ = 39.54 K, $A_{ul}$ = 3.45~10$^{-4}$ s$^{-1}$) lies at the same frequency (246209.0970 MHz). However, no other transitions with similar $E_{up}$ and $A_{ul}$ were detected in the continuum {\it spw} within the noise. No conclusion can be determined so far.

\subsection{Non-detected iCOMs}
\paragraph{Acetaldehyde}
Acetaldehyde (CH$_{3}$CHO) was not detected in our observations. This might be surprising since it is routinely  detected in hot corinos \citep[][e.g.]{Cazaux2003,Jorgensen2016,Bianchi2022} and protostellar shocks \citep[][e.g.]{Arce2008,Lefloch2017,DeSimone2020-I4outflow}. It is the most abundant iCOM after methanol \citep[e.g.][]{Ceccarelli2023}. Twenty potential transitions are spread over the two continuum bands and are listed in Table \ref{tab: spectroND} in the appendix. However, the most likely line candidate for detection in these bands has a low Einstein coefficient (6.6~$\times$ 10$^{-5}$ s$^{-1}$), a relatively high upper energy of 96 K, and is blended with a bright \chtocho~transition ($A_{ul}$ = 2 $\times$ 10$^{-4}$ s$^{-1}$). 
\paragraph{Nitrogen-bearing iCOMs}
We also note that ethyl cyanide (C$_2$H$_5$CN) and vinyl cyanide (C$_2$H$_3$CN), which are often observed in high-mass star-forming regions \citep{Fontani2007}, were not detected. Within the observed frequency range, the C$_2$H$_5$CN transition at 246.2687 GHz and the C$_2$H$_3$CN transition at 247.0870 GHz are expected to be brightest, but neither is detected (see Table \ref{tab: spectroND} in the appendix).

\subsection{Other lines}
Finally, in the two continuum windows (see Table \ref{tab: spw}), we identified and detected the \chtoh~ transitions published by \citet{Vastel2022}, one HDCO transition published by \citet{Evans2023}, two \sot~transitions, as well as one SO and S$^{34}$O transition (to be published in a subsequent FAUST publication).

\section{Spatial distribution}
\label{sec: results}

\begin{figure}
    \centering
    \includegraphics[width = 0.48\textwidth,trim=0 0 0 0 clip]{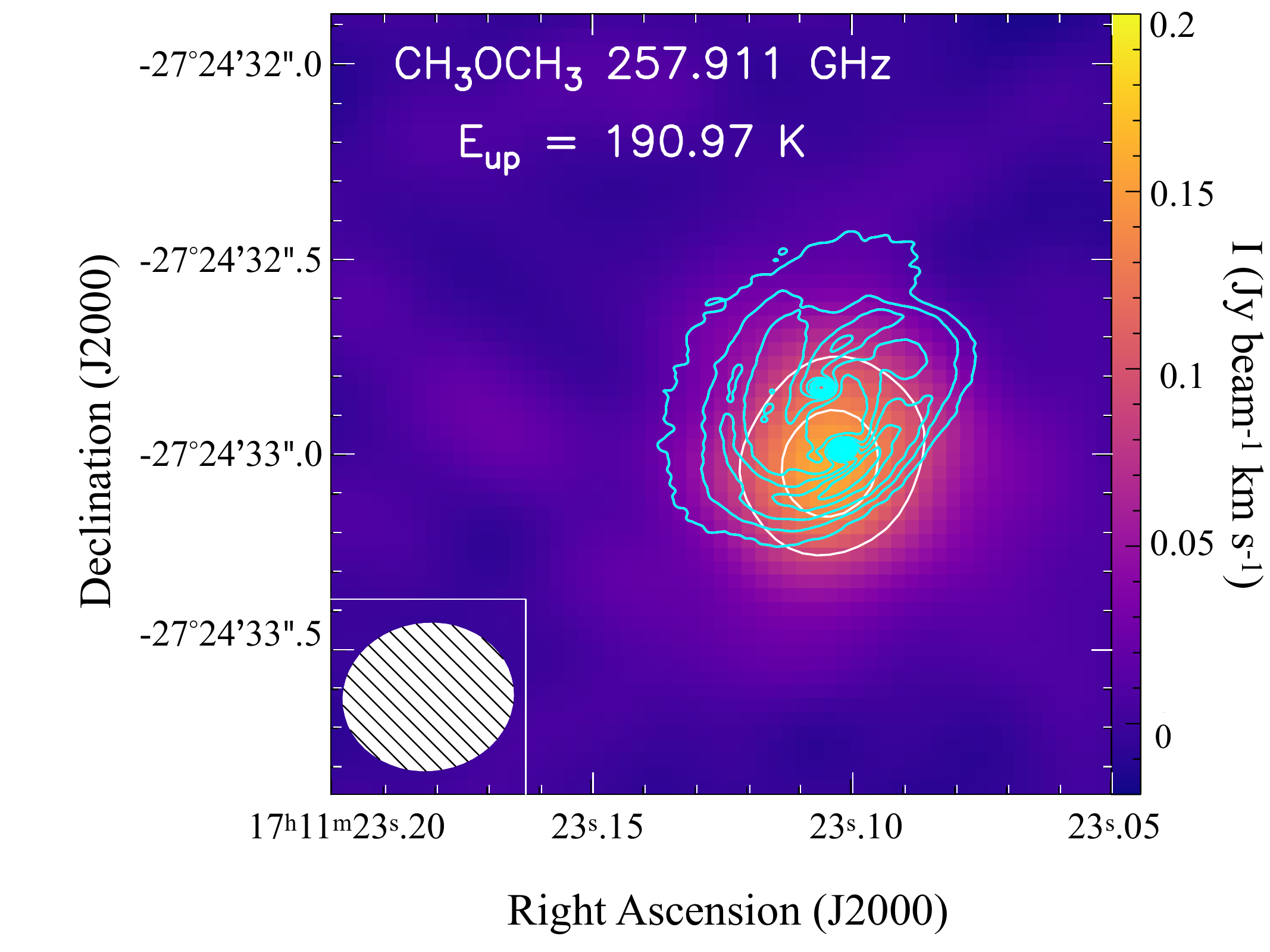}
    \includegraphics[width=0.48\textwidth,trim=0 0 0 0 clip]{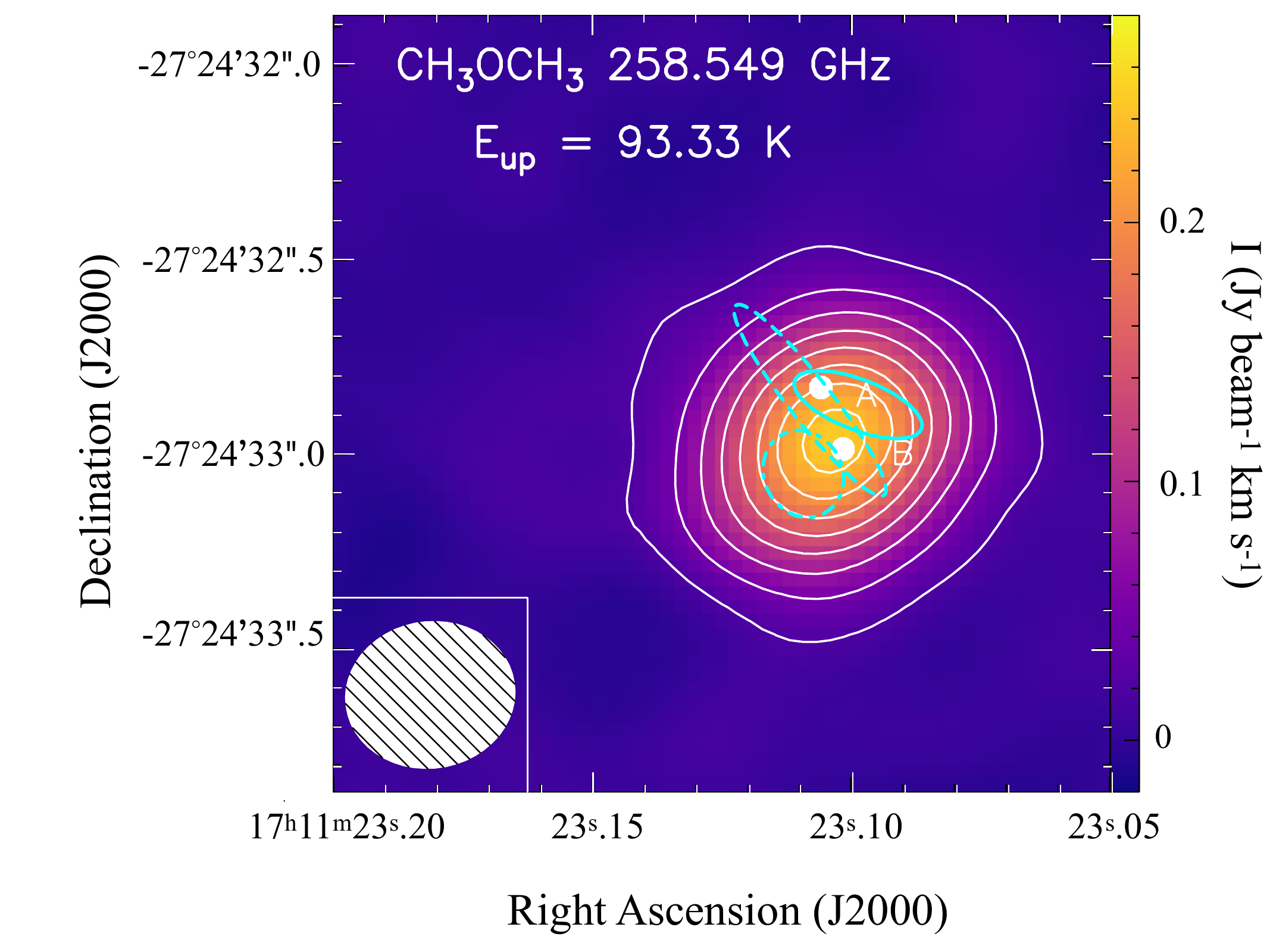} \ 
    \caption{Moment 0 maps for the two high spectral resolution \chtocht~transitions. {\it Top:} The 257.911 GHz transition. The contours in white start at 4$\sigma$ at every 2$\sigma$ for 257.911 GHz (1$\sigma$ =  22.5 mJy~beam$^{-1}$~km~s$^{-1}$). The dust continuum map from \citet{Alves2019} is superposed in cyan. {\it Bottom:} The 258.549 GHz transition. The contours in white start at $\sigma$ every 6$\sigma$ at 258.548 GHz (1$\sigma$ =  5.1 mJy~beam$^{-1}$~km~s$^{-1}$). The image component sizes deconvolved from the beam are indicated with an ellipse (in cyan) at the -2 km~s$^{-1}$ channel, a dashed ellipse at the 2.8~km~s$^{-1}$ channel, and a dot-dashed ellipse at the 9.9~km~s$^{-1}$ channel. Sources A and B are indicated as filled white circles. In both figures, the ellipse in the bottom left corner represents the ALMA synthesised beam. }
    \label{fig: mom0-ch3och3}
\end{figure}
\begin{figure}
    \centering
    \includegraphics[width=0.45\textwidth]{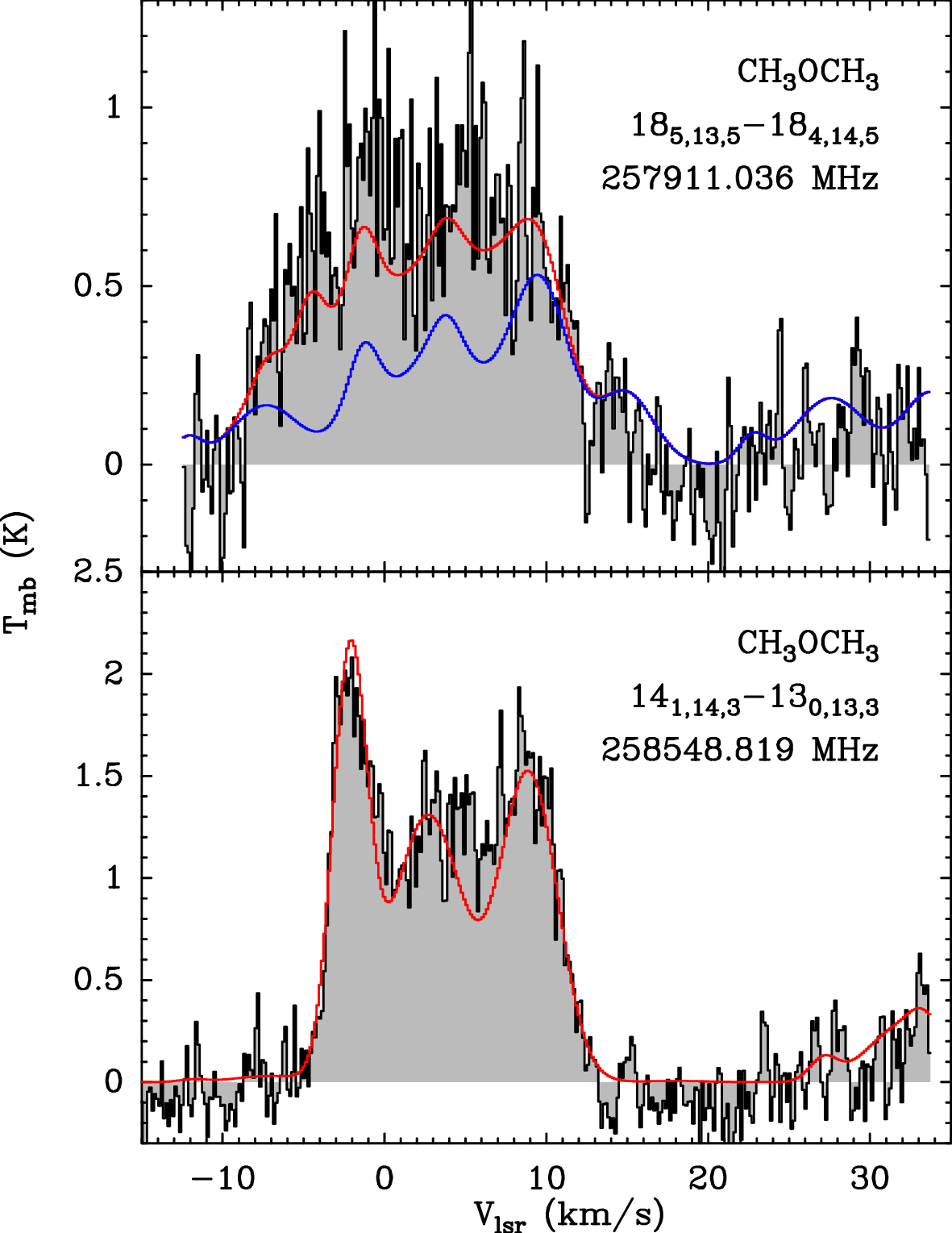}
    \caption{Observed transitions of \chtocht~(in black). The best-fit model is superposed in red (corresponding to the LTE modelling of both \chtocht~and \chtocho~ explained in Sect. \ref{sec: modelling}), and the corresponding spectroscopic parameters are listed in Table \ref{tab: spectro}. The transition at 257.911 GHz overlaps with a \chtocho~transition whose best-fit model (based on the many lines detected in our observations) is shown in blue. The grey shaded area corresponds to the filled histogram above and below the baseline.}
    \label{fig: spec-ch3och3}
\end{figure}
\begin{figure*}[ht!]
    \centering
    \includegraphics[width=0.46\textwidth]{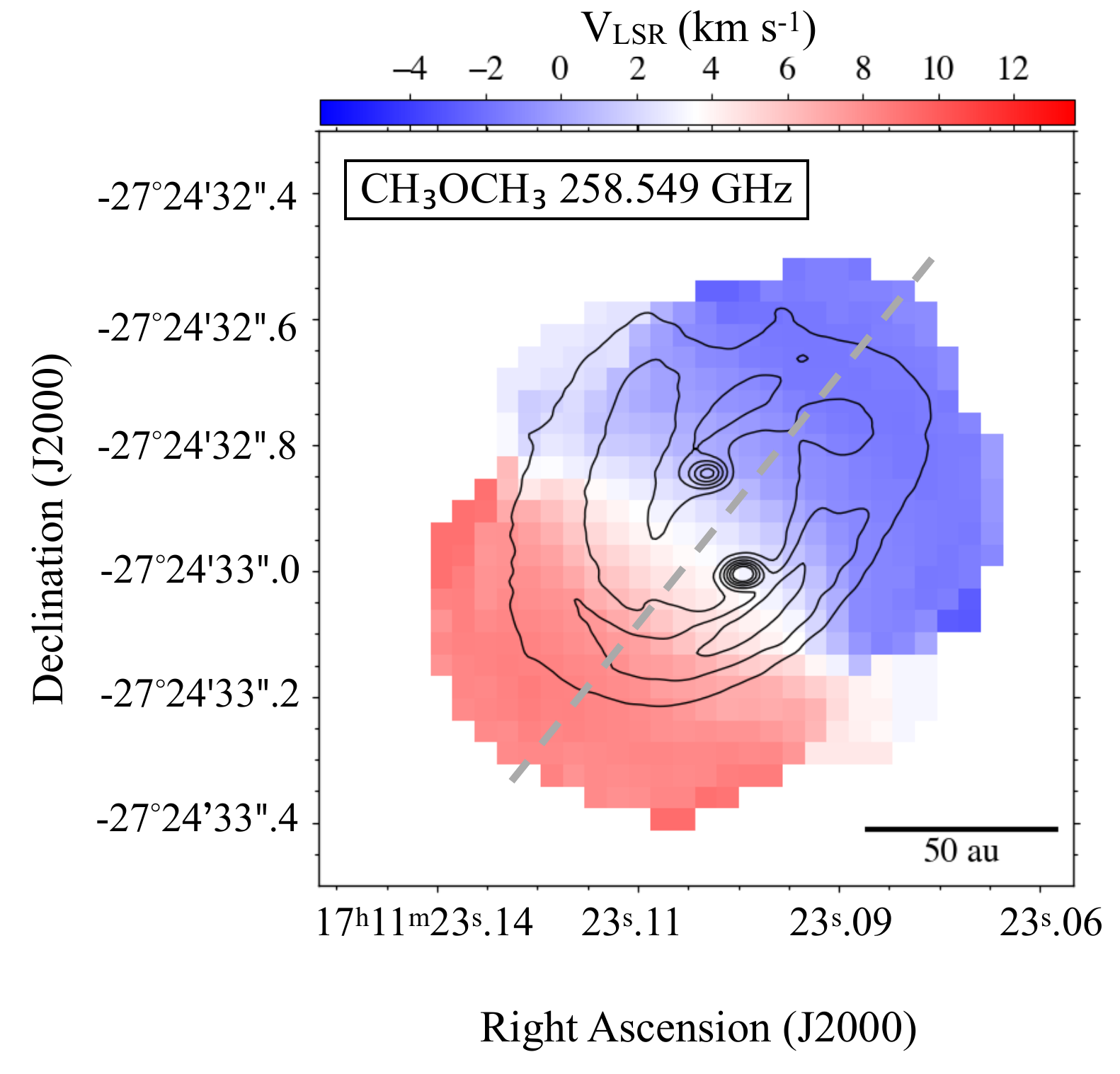}
     \includegraphics[width=0.36\textwidth]{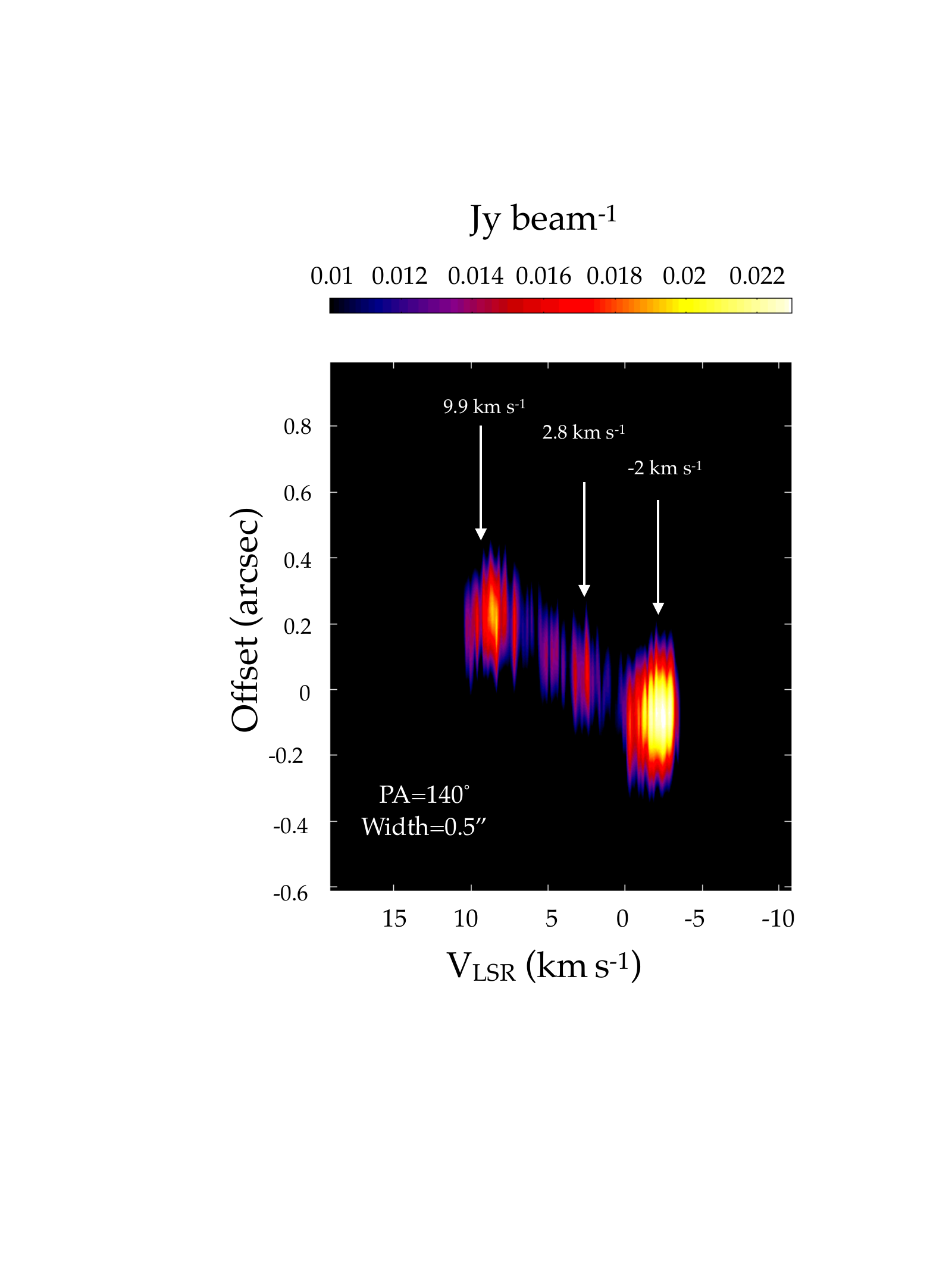}
    \caption{Velocity analysis based on the \chtocht~ transition at 258.549 GHz. {\it Left:} The contours for the velocity field indicate the continuum intensity levels from the high spatial resolution observations of \citet{Alves2019}. The compact emission contours arise from the circumstellar disks around each protostar of the binary system, 11A (northern component) and 11B (southern component). The dashed line shows the direction of the cut used to produce the PV diagram displayed in Fig. \ref{fig: moment1-ch3och3} centred on a position in between the two cores. The cut position angle is 140$^{\circ}$ and its width is 0.5$^{\prime\prime}$, encompassing both protostars. {\it Right:} PV diagram obtained from the \chtocht~emission at 258.549 GHz. The direction of the cut (PA) and its width used to produce the PV diagram is shown on the left side of the figure as a grey dashed line. The three velocity components observed in the methanol spectra are also indicated \citep{Vastel2022}.}
    \label{fig: moment1-ch3och3}
\end{figure*}

The maps presented in the following were obtained using the IRAM/GILDAS\footnote{\url{http://www.iram.fr/IRAMFR/GILDAS}} package (version dec22a). Figure \ref{fig: mom0-ch3och3} shows the moment 0 maps for the two strong \chtocht~transitions. The spectra from both transitions were extracted from a $\sim$ 0.4$^{\prime\prime}$ beam size, corresponding to the beam for both transitions (see Table \ref{tab: spectro}), which includes the 11A and 11B cores. This corresponds to $\sim$ 30$\sigma$ for the 258.5 GHz transition (upper energy level of 93 K) and also corresponds to $\sim$ 4$\sigma$ for the 257.9 GHz emission (upper energy level at 191 K). The line profiles are presented in Fig. \ref{fig: spec-ch3och3}. The 257.911 GHz line profile is more widespread in frequency and does not present a simple three-component feature because of some blending with \chtocho~transitions at 257.9 GHz (see Table \ref{tab: spectro}, $\rm E_{up}$ = 306.01 K). The 258.549 GHz transition of dimethyl ether is the strongest line of the detected iCOMs presented in this study that is not affected by blending from other transitions. The profile is a simple three-fit Gaussian, and we used it to try to determine the exact location, assuming that the emission is compact. In Fig. \ref{fig: mom0-ch3och3} (bottom), the image component size deconvolved from the beam for the central component (2.8~km~s$^{-1}$) is more extended than the other, and we cannot yet deduce a clear location (see more details in Sec. \ref{sec: discussion}). Furthermore, we used the unblended 258.549 GHz transition of dimethyl ether to compute the velocity field (moment 1) around \bhb. The left side of Fig. \ref{fig: moment1-ch3och3} shows a clear northwest-southeast gradient centred in between 11A and 11B at the systemic velocity of the source. The same velocity field has been obtained for the methanol transition \citep{Vastel2022}.

In order to better understand the left side of Fig.  \ref{fig: moment1-ch3och3}, we created a position-velocity (PV) diagram for the same transition. The PV plot shown on the right side of the figure was produced from a cut oriented parallel to the velocity gradient observed in the moment 1 map of the dimethyl ether emission at 258.549 GHz, whose PA is 140$^{\circ}$ east of north. The cut width is 0.5$^{\prime\prime}$, comparable to the angular resolution of the FAUST data, and it encompasses cores 11A and 11B, which are separated by 0.2$^{\prime\prime}$. This figure clearly shows the three velocity structures that were identified with methanol \citep{Vastel2022} and does not show a smooth Keplerian profile over the velocity range from the circumbinary envelope, as was shown using formaldehyde by \citet{Alves2019}. 

\begin{figure*}
    \centering
    \includegraphics[width=0.44\textwidth]{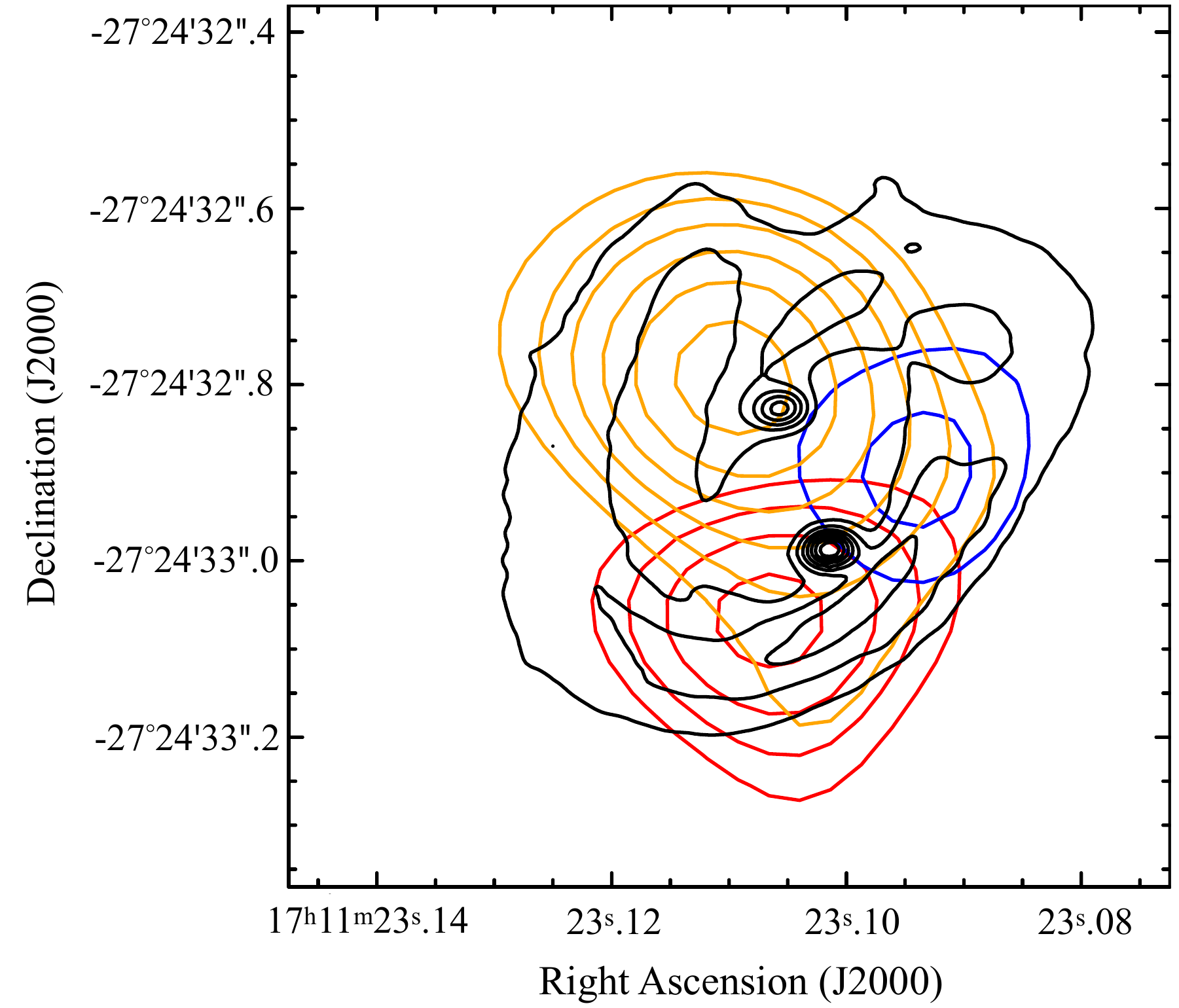}
     \includegraphics[width=0.55\textwidth]{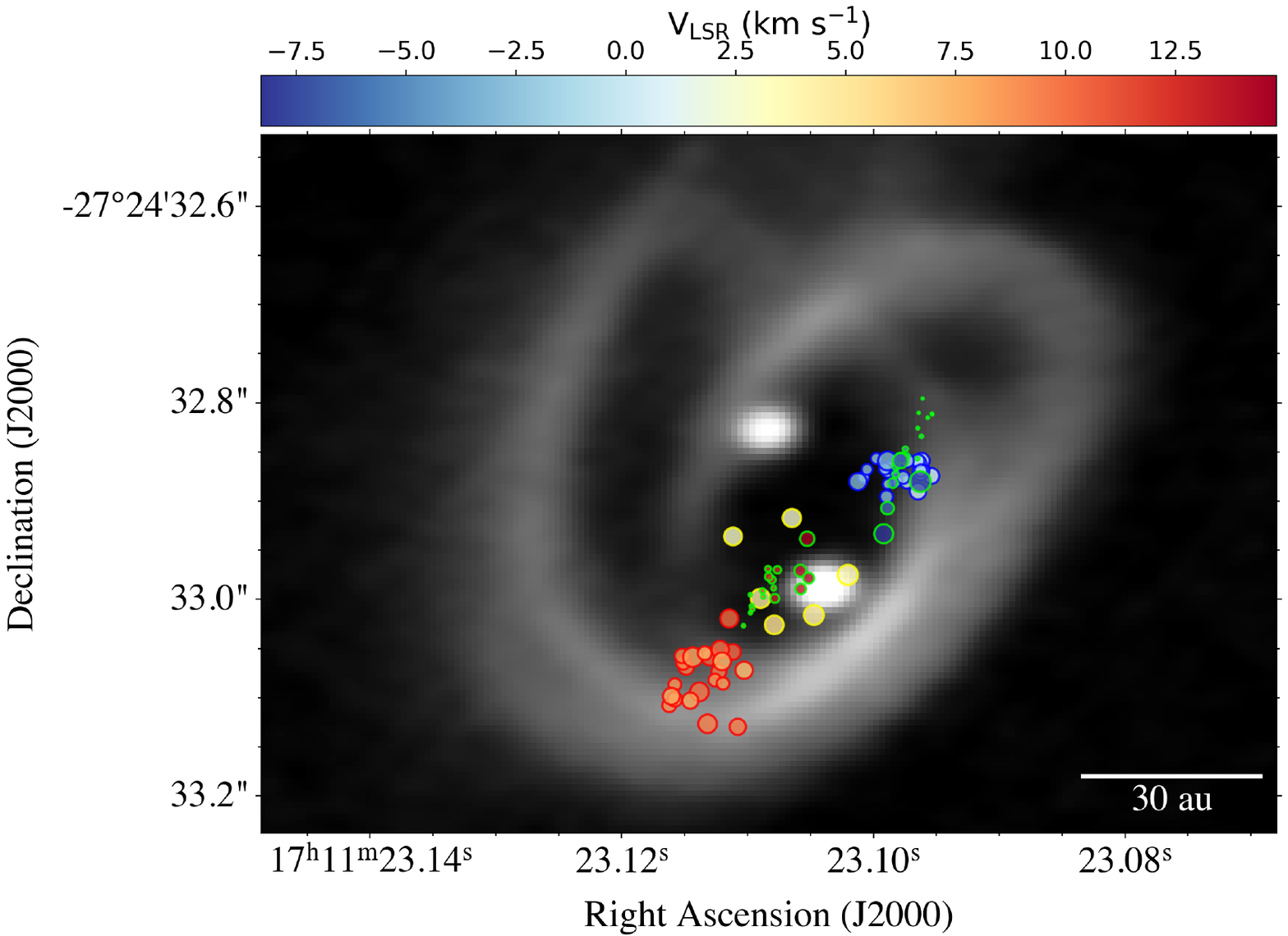}
    \caption{Analysis of the origin of the observed dimethyl ether components. {\it Left:} Contour map emission at 258.549 GHz for the two most extreme velocities detected at 3$\sigma$ at the -3.69 km~s$^{-1}$ channel (blue) and at the 11.86 km~s$^{-1}$ channel (red), as well as the central peak channel at 2.8 km~s$^{-1}$ (orange). The contours start at 4$\sigma$ every 1$\sigma$. The dust continuum map from \citet{Alves2019} is superposed in black. {\it Right:} Position of intensity peaks (circles) from the 2D Gaussian fit of the emission of the \chtocht~line transition at 258.549 GHz in each velocity channel separated by $\sim 0.14$ km s$^{-1}$ plotted over the dust continuum map of \citet{Alves2019}. The colour code for the edges of the circle is red for the 9.2 km~s$^{-1}$ \chtocht~component ($\ge$ 6 km~s$^{-1}$), yellow for the 3 km~s$^{-1}$ \chtocht~component ($]1, 6[$ km~s$^{-1}$ ), and blue for the -1.9 km~s$^{-1}$ \chtocht~component ($< $1 km~s$^{-1}$), while the CO (2--1) high-velocity components are shown in bright green. The circle size is defined by the uncertainty on the absolute position of the emission peak in each velocity component (the largest circle has $\sim 0.02^{\prime\prime}$ given by $\theta$/(2xS/N), as explained in the text of Sect. \ref{sec: results}).}
    \label{fig: 3comp-ch3och3}
\end{figure*}

Considering the angular resolution of the present observations, we cannot determine the exact location of the iCOMs emission within the binary system. However, we applied the same method as in \citet{Vastel2022} and present on the left side of Fig. \ref{fig: 3comp-ch3och3} the red- and blueshifted emission for \chtocht~in the high-velocity channels where the S/N is higher than five (the channel at -3.69 km~s$^{-1}$ in the blue and the channel at 11.18 km~s$^{-1}$ in the red). Both components tend to be aligned with source 11B, like the high-velocity components of the CO emission \citep{Alves2019} and the methanol emission \citep{Vastel2022}. The emission in the 2.8 km~s$^{-1}$ channel (yellow) peaks near 11A in the same way as CO and \chtoh. The three \chtocht~components appear in a compact configuration that is compatible with the compact emission seen in the moment 0 map (see Fig. \ref{fig: mom0-ch3och3}).

Since the iCOMs emission in each velocity channel is compact and centrally peaked, we performed 2D Gaussian fits on the \chtocht~transition at 258.549 GHz (the strongest unblended transition with a high S/N) in each channel to obtain the position of their intensity peak in the three components [-5, 1] km~s$^{-1}$, [1, 6] km~s$^{-1}$, and [6, 13] km~s$^{-1}$. The right side of Fig. \ref{fig: 3comp-ch3och3} shows this 2D analysis. The accuracy on the relative position is given by $\theta/(2 \times S/N)$, where $\theta$ is the synthesised beam size, and S/N is the signal-to-noise ratio of the peak with respect to the map noise \citep{Condon1997,Condon1998}. Consequently, the uncertainty on the position will be lower for higher S/N line intensities. In other words, a centroid fitting to the compact molecular emission yields astrometry with a precision better than that of the spatial resolution in the case of a high S/N. For this calculation, we only considered intensity peaks brighter than five times its Gaussian fit error and a S/N higher than ten, corresponding to a position accuracy better than 0.02$^{\prime\prime}$. This 2D analysis is similar to our analysis of methanol, except for the middle component at $\sim$ 3 km~s$^{-1}$, which is more localised on the B source instead of in between the two cores. We note, however, the limited number of channels satisfying the requirements (S/N $>$ 10, intensity/error $>$ 5) compared to methanol. Fig. \ref{fig: 3comp-ch3och3} (right) shows no clear acceleration of the iCOMs material towards \bhb~B (circles with red edges for velocities greater than 6 km~s$^{-1}$ and circles with blue edges for velocities lower than 1 km~s$^{-1}$), in contrast to what is seen in the high-velocity component of CO (circles with green edges) with a darker colour inside the circles nearby the source.

\begin{figure}
    \centering
    \includegraphics[width = 0.45\textwidth,trim=0 0 0 0 clip]{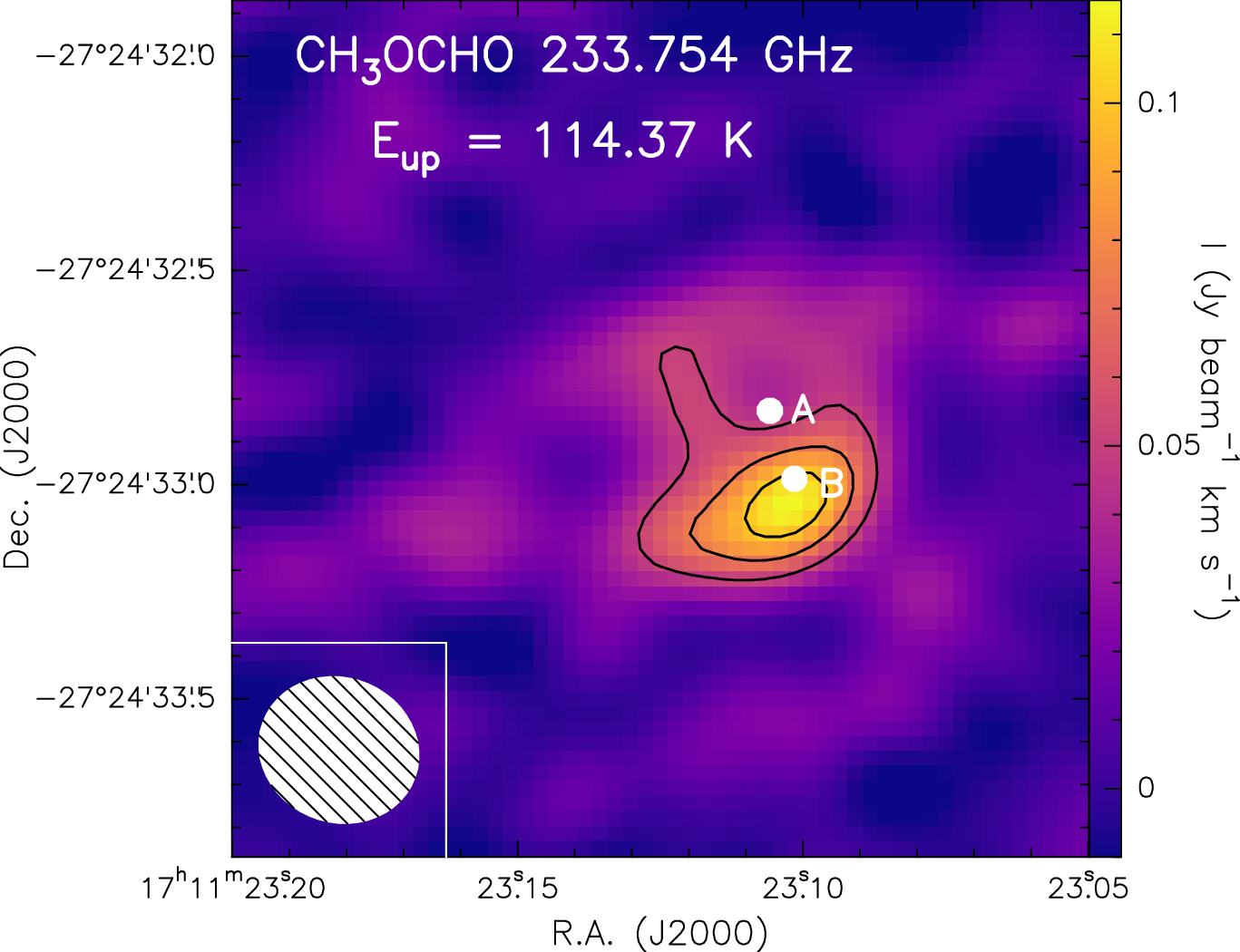}
    \includegraphics[width=0.45\textwidth,trim=0 0 0 0 clip]{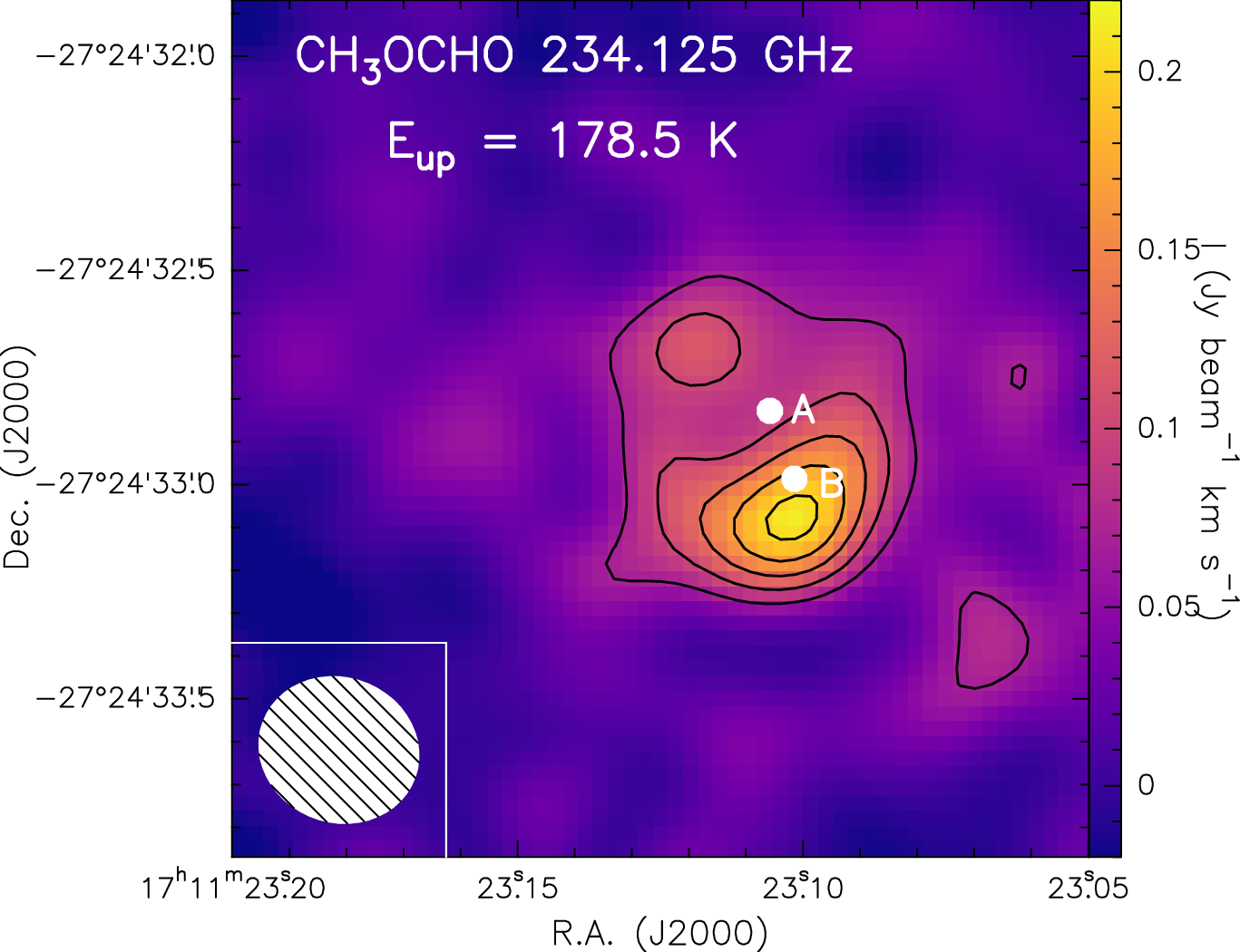} 
        \includegraphics[width=0.45\textwidth,trim=0 0 0 0 clip]{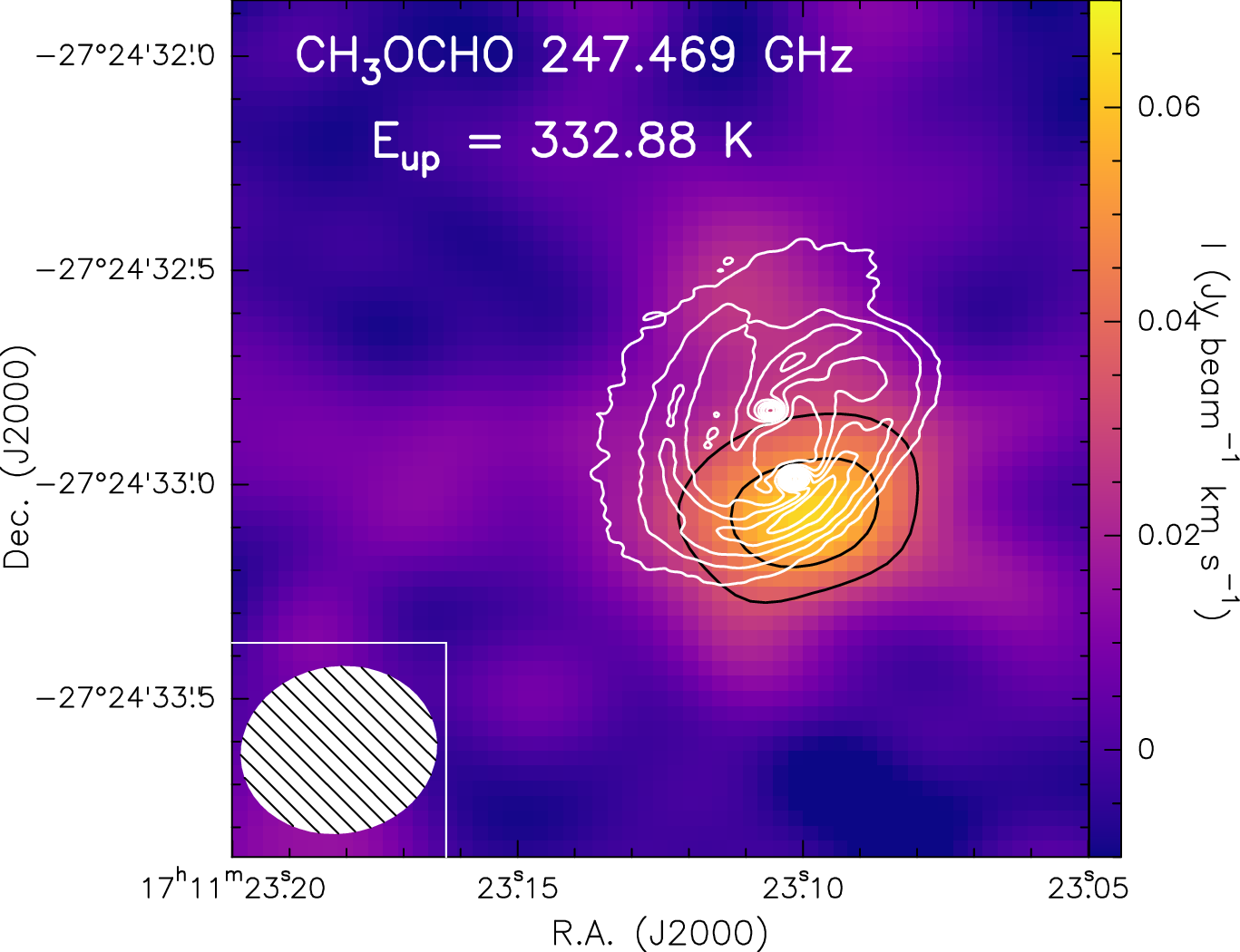} \
    \caption{Moment 0 maps for some \chtocho~ transitions. The contours start at 4$\sigma$ at every 2$\sigma$ (1$\sigma$ =  11.7, 16.7, and 8.3  mJy~beam$^{-1}$~km~s$^{-1}$ at 233.8, 234.1, and 247.5 GHz, respectively). The ellipses in the bottom left corners represent the ALMA synthesised beams. Sources A and B are indicated as filled white circles in the top and middle panels. The dust continuum map from \citet{Alves2019} is shown in white contours in the bottom panel.}
    \label{fig: mom0-ch3ocho}
\end{figure}

\begin{figure*}
    \centering
    \includegraphics[width=0.8\textwidth]{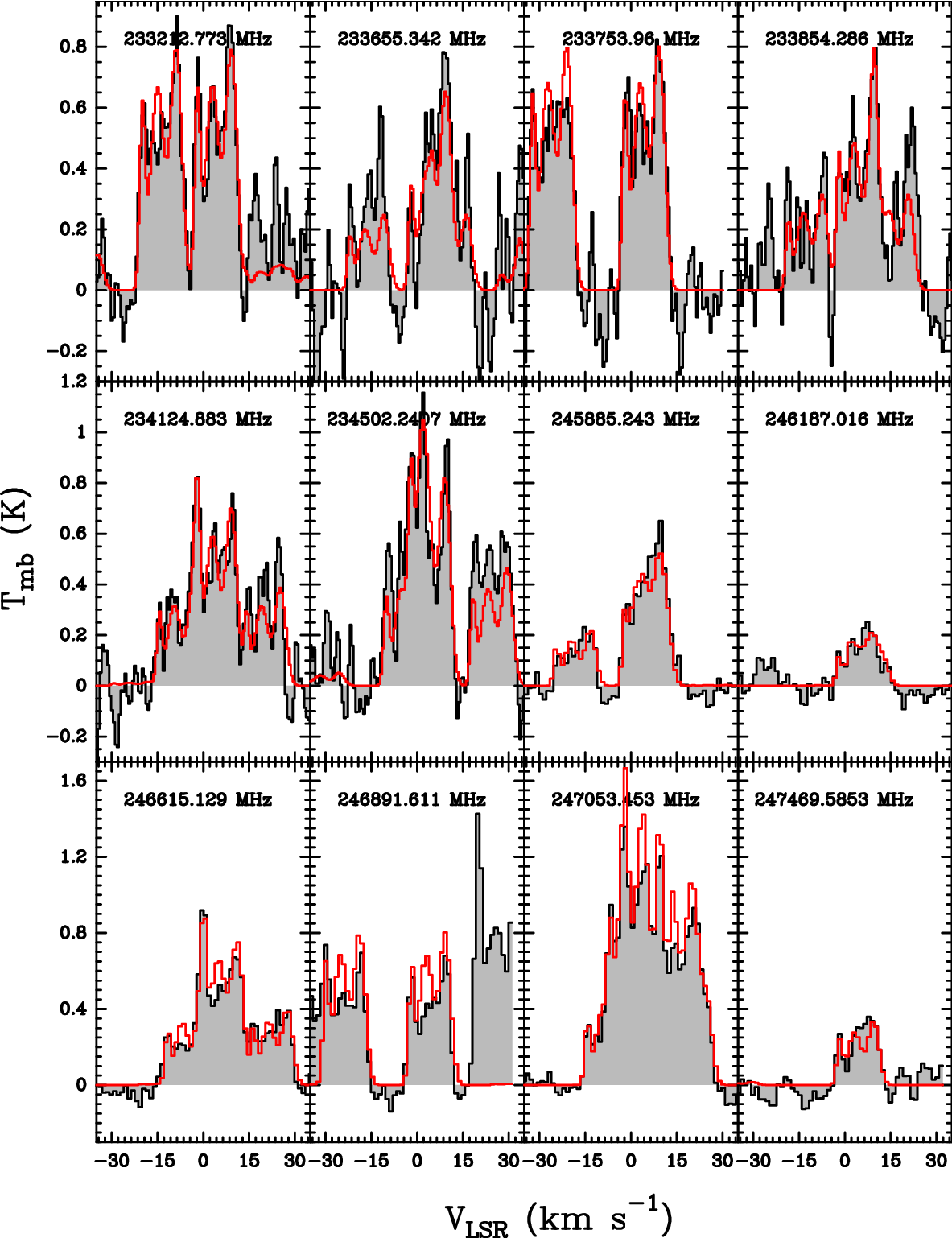}
    \caption{Example of isolated \chtocho~transitions observed within 0.8$^{\prime\prime}$ (in black). The best-fit model is superposed in red, and the corresponding spectroscopic parameters are listed in Table \ref{tab: spectro}. The corresponding $V_{LSR}$ for the three components are -1.9, 3.0, and 9.2 km~s$^{-1}$ , and for the FWHM, they are 2, 5, 4 km~s$^{-1}$, respectively. The grey shaded area corresponds to the filled histogram above and below the baseline.}
    \label{fig: spec-ch3ocho}
\end{figure*}

Many transitions of methyl formate (\chtocho) were detected in the continuum band (Fig. \ref{fig: cont-spectra}), and Fig. \ref{fig: mom0-ch3ocho} shows the moment 0 maps for three isolated transitions covering a wide upper energy range between 115 and 333 K. 
The smaller beam of the transition at 233 GHz shows that the emission peaks south of \bhb~B appear to clearly show a lack of emission towards \bhb~A. There is a good correlation between this higher intensity in the southern region with the kinematic structure in the form of a clear velocity gradient around source \bhb~B (see Fig. \ref{fig: 3comp-ch3och3}, right). Figure \ref{fig: spec-ch3ocho} shows the spectra of some isolated transitions of \chtocho~in the continuum {\it spw}. \\

\begin{figure*}
    \centering
    \includegraphics[width=0.47\textwidth]{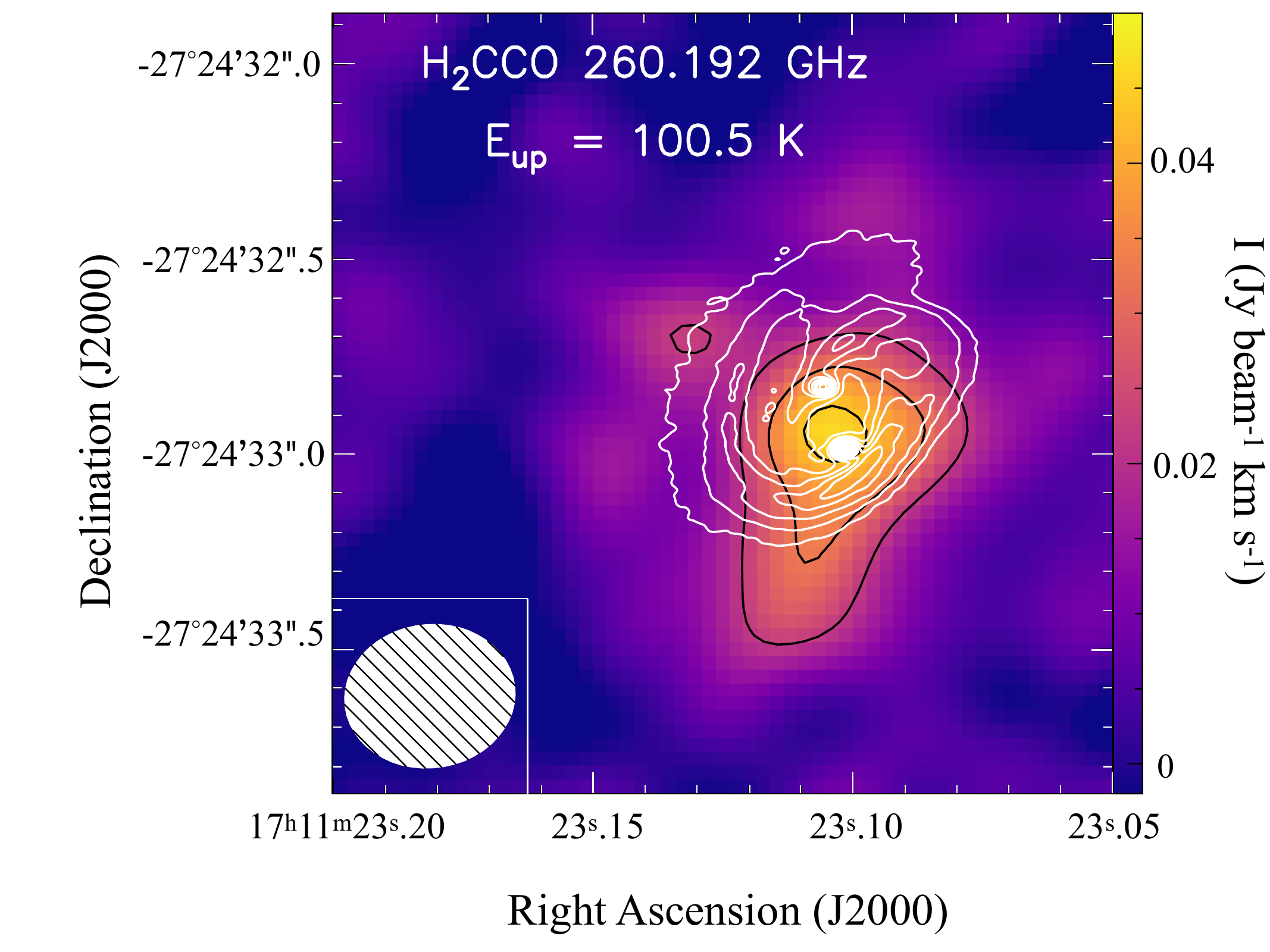}
     \includegraphics[width=0.45\textwidth]{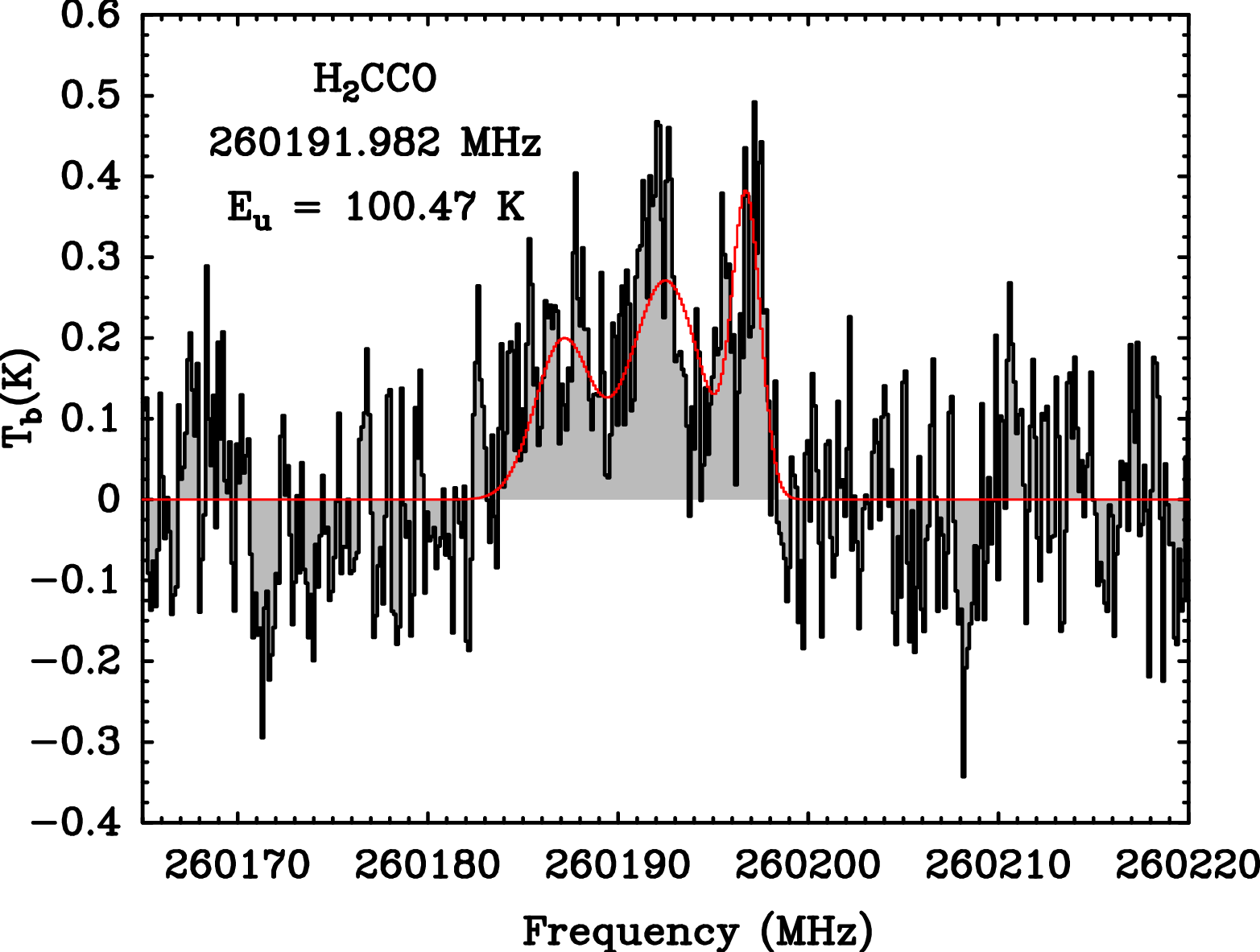}
    \caption{Analysis of the observed ketene transition. {\it Left:} Moment 0 map at 260.192 GHz. The contours start at 4$\sigma$ at every 2$\sigma$. The ellipse in the bottom left corner represents the ALMA synthesised beam ($0.44^{\prime\prime} \times 0.37^{\prime\prime}$). Sources A and B identified by \citet{Alves2019} are indicated as filled white circles. {\it Right:} H$_2$CCO transition at 260.192 GHz. The red line corresponds to the modelled transitions as described in Sect. \ref{sec: modelling}. The grey shaded area corresponds to the filled histogram above and below the baseline.} 
    \label{fig: spec-h2cco}
\end{figure*}

The left side of Fig. \ref{fig: spec-h2cco} shows the moment 0 map for the only transition of ketene, again, with an enhancement in intensity in the southern region of 11B, following the filament traced with the dust continuum emission presented by \citet{Alves2019}, as shown in Fig. \ref{fig: 3comp-ch3och3} (right).

\section{Radiative transfer modelling}
\label{sec: modelling}

We first performed a local thermodynamic equilibrium (LTE) analysis to estimate the \chtocht/\chtoh~and \chtocho/\chtoh~abundance ratios. The methanol column density was computed from non-LTE modelling using its $^{13}$C isotopologue \citep{Vastel2022}. For methyl formate and dimethyl ether, we extracted the spectra from the same region as for the \chtocht~ transition at 257.911 GHz, corresponding to $\sim$ 0.4$^{\prime\prime}$. 
We then used the parameters (column density, source size, and kinetic temperatures) found from the non-LTE modelling of the methanol transitions \citep[see][]{Vastel2022} to compute the column densities of \chtocht~and \chtocho~using the CASSIS software. We fixed the source sizes and let the best-fit model converge around the $V_{LSR}$  line width for the Gaussian fits and the kinetic temperatures of methanol for the excitation temperatures of the three components. We then let the source size vary. Table \ref{tab: LTE} lists the results from the best fit obtained using all transitions in both continuum windows and in setup 2, {\it spw} 47 and 43, taking into account the blended lines of \chtocht~ and \chtocho. Figures \ref{fig: spec-ch3och3} and  \ref{fig: spec-ch3ocho} show the best-fit LTE model (in red) obtained by simultaneously fitting all lines of \chtocht~ and \chtocho. The corresponding $V_{LSR}$ for the three components are -1.9, 3.0, and 9.2 km~s$^{-1}$, and the FWHM values are 2, 5, and 4 km~s$^{-1}$ respectively.\\
With methanol as a reference, the \chtocht/\chtoh~ratio is [5, 8] $\times$ 10$^{-3}$ and the \chtocho/\chtoh~ratio is   [2.5, 4.5] $\times$ 10$^{-2}$ for the three components.\\

\begin{table*}
\caption{Results from the LTE analysis based on the non-LTE results from methanol. }
\label{tab: LTE}
\centering
\begin{tabular}{c c c c }
\hline 
\hline
 & &Velocity components & \\
\hline 
\hline
                                                     &  -2 km~s$^{-1}$                                  & 2.8 km~s$^{-1}$ & 9.9 km~s$^{-1}$   \\
\hline \hline
N(\chtoh) (cm$^{-2}$)        & ($1.4 \pm 0.6$) $\times 10^{18}$      &  ($4.0 \pm 1.0$) $\times 10^{18}$  & ($3.0 \pm 0.8$) $\times 10^{18}$ \\
T (K)                                             &  $110 \pm 10$                                   & $130 \pm 10$   & $130 \pm 10$ \\
n$_{H_{2}}$ (cm$^{-3}$)              & ($1.0^{+3}_{-0.5}$) $\times 10^{7}$        &($2.0 \pm 0.5$) $\times 10^{6}$ & ($2.5 \pm 2.2$) $\times 10^{7}$\\
$\theta_{S}$ ($^{\prime\prime}$)  &  $0.15^{\prime\prime} \pm 0.10^{\prime\prime}$ & $0.13^{\prime\prime} \pm 0.10^{\prime\prime}$  & 0.12$^{\prime\prime} \pm 0.10^{\prime\prime}$\\
\hline\hline
N(\chtocho)  (cm$^{-2}$)     &  [1.0, 3.4] $\times 10^{16}$   & [2.2, 10.0] $\times 10^{16}$  & [3.0, 13.0] $\times 10^{16}$\\
N(\chtocht)  (cm$^{-2}$)      &  [3.5, 11.0] $\times 10^{15}$   &  [6, 20] $\times 10^{15}$  &  [6, 25] $\times 10^{15}$\\
N(t-HCOOH)  (cm$^{-2}$)   &  [0.6, 2.5] $\times 10^{15}$   & [0.55, 1.5] $\times 10^{15}$  & [1.0, 3.5] $\times 10^{15}$\\
N(H$_{2}$CCO)  (cm$^{-2}$)   & [2.5, 20] $\times 10^{14}$   & [6, 25] $\times 10^{14}$  & [4, 20] $\times 10^{14}$\\
$\theta_{S}$ ($^{\prime\prime}$)  & [0.15, 0.30] & [0.13, 0.26]  & [0.12, 0.24]\\
T$_{ex}$   (K)                           &    [80, 100]                             & [80, 100]                 & [80, 100]  \\
\hline
\end{tabular}
\tablefoot{The methanol non-LTE analysis and the three velocity components identifications are from \citet{Vastel2022}.}
\end{table*}

We detected a line at 246.79 GHz (see Fig. \ref{fig: ch3o13cho}) that may correspond to the blending of four 13C$_{2}$ methyl formate (CH$_{3}$O$^{13}$CHO) transitions. The spectroscopic parameters are listed in Table \ref{tab: spectroND}. Many lines have previously been detected in the spectrum of the Orion star-forming region \citep{Carvajal2009} at the IRc2 position. Combining our \chtocho~column density with its isotopologue, we used a $^{12}$C/$^{13}$C ratio of 68 \citep{Milam2005} to compute a CH$_{3}$O$^{13}$CHO ratio of [1.5, 5.0] $\times$ 10$^{14}$, [3.2, 15] $\times$ 10$^{14}$ and [4.4, 19] $\times$ 10$^{14}$ cm$^{-2}$. With the same source size and excitation temperatures as for \chtocho~and \chtocht, the modelled transitions using an LTE radiative transfer analysis in the two continuum bands are compatible with the observations within the noise. However, the LTE spectrum (Fig. \ref{fig: ch3o13cho}, at the frequency of the middle group of 246788.406 MHz) presents a shift in velocity for the three components that might be explained by an unidentified transition at this position that is not reported in the CDMS and JPL databases, or by an incorrect frequency in the CH$_{3}$O$^{13}$CHO database. The quoted uncertainty of the database frequency is only 2 kHz, which cannot explain this shift. The 246.788 GHz line also appears in the Orion spectra \citep[Fig. 2 of][]{Carvajal2009}, and it is well underestimated by the model with a clear shift as well. However, the authors did not explain this particular shift. Considering the noise and the spectral resolution of our observations for this transition, we cannot conclude on a firm detection of the methyl formate isotopologue.

We used the CH$_{3}$OCHO and CH$_{3}$OCH$_{3}$ LTE modelling to obtain a good estimate of the formic acid column density. The 246.106 and 247.514 GHz lines are blended with methyl formate and are underestimated by the LTE model when formic acid is not taken into account. Taking the three components with the same beam size, excitation temperature, and $V_{LSR}$ as derived from the methyl formate and dimethyl ether into account, we obtained the column densities quoted in Table \ref{tab: LTE}. The 246.106 and 247.514 GHz lines are now better reproduced by the LTE model. When methanol is again used as a reference, we obtain a t-HCOOH/\chtoh~ ratio of [0.01, 0.5] $\%$. \\

We present on the right side of Fig. \ref{fig: spec-h2cco} the only transition for ketene, as discussed in Sect. \ref{sec: line-id}. The column densities can be constrained from the LTE modelling, again using the excitation temperature, filling factor, and FWHM as for \chtocht~and \chtocho. The results are presented in Table \ref{tab: LTE} with a H$_2$CCO/\chtoh~ratio of [0.01, 0.1] $\%$, which is very similar to the t-HCOOH/\chtoh~ratio.\\

An upper limit for acetaldehyde column density can be estimated at about 1/70th compared to the methanol column density. In Fig. \ref{fig: spec-ch3ocho}, the transition at 234.486 GHz (see Table \ref{tab: spectro}) is not well reproduced. However, it  might be contaminated by an acetaldehyde transition at 234.486 GHz (see Table \ref{tab: spectroND}), which is within the upper limit value calculated above (0.15 K for this transition for a 0.115 K $rms$). We also evaluated the upper limit of the C$_2$H$_5$CN and C$_2$H$_3$CN column densities for each velocity component with the LTE analysis. The derivation of the upper limits is described in the appendix. We obtained C$_2$H$_5$CN/CH$_3$OH and C$_2$H$_3$CN/CH$_3$OH to be $<$0.066 \% and $<$0.018 \%, respectively.

\section{Discussion}
\label{sec: discussion}

\subsection{Origin of the emission of iCOMs in \bhb}

The spatial distribution of iCOMs in hot corinos around Class 0 protostars is known to be compact and to peak at the location of the dust continuum peak, which corresponds to the location of the central source \citep[see e.g.][]{LopezSepulcre2017,Jorgensen2018}. In contrast, \bhb~shows hot methanol gas that is not exactly associated with the location of the central protostars of the binary system, 11A and 11B, but distributed across the circumbinary disk \citep{Vastel2022}. Even though the binary system cannot be resolved in the FAUST images, the 2D Gaussian centroid map of the hot methanol line at 243.916 GHz reveals that this emission appears to be slightly offset with respect to 11A ($\sim$ 20 au) and towards the north-west and south-east of 11B ($\sim$ 20 au from the hot core), at the intersection between filaments seen in the dust emission. The non-LTE analysis of the methanol lines tells us that the gas is hot ($\sim$ 120 K)  and dense [0.2, 25] $\times$ 10$^{7}$ cm$^{-3}$ \citep{Vastel2022}. The dust in between 11A and 11B could be warm enough for the methanol to sublimate, but the iCOM emission does not peak on the hot cores themselves, which seems to rule this hypothesis out. 
Alternatively, the hot methanol gas could be associated with shocked gas. From the theoretical point of view, shocks may occur in mass-transferring binary systems, where spiral shocks are induced by disturbances and accretion within the disk \citep{Spruit1987,Ju2016}. They also occur as streaming gas from the circumbinary disk to the circumstellar disks \citep{Mosta2019} and impact streams close to the inner edge of the circumbinary disk \citep{Shi2012}. 

In \citet{Vastel2022}, we proposed that hot methanol arises from shocked gas associated with an accretion gas streamer falling onto source 11B, which induces a shock interaction that is strong enough to inject species from the ice mantles in the gas phase. The impact is sufficient to overcome the binding energies, and this mechanism is called sputtering \citep[e.g.][]{Draine1995}. \citet{Alves2019} already suggested that the filaments surrounding \bhb~are accretion streamers from the extended circumbinary disk onto the circumstellar disks of the protobinary system, from large- to small-scale structures. This streaming material would impact either on the quiescent gas of the circumbinary envelope or possibly on the two circumstellar disks. Recently, the impact of these accreting streamers on the gas chemical enrichment has been observed in some very few sources, in addition to the indirect case of \bhb~: the Class II sources DG Tau and HL Tau  \citep{Garufi2022}, and the Class I source SVS13 A \citep{Bianchi2023,Hsieh2023}. In addition, only a few simple species have been detected in these streamer-triggered shocks so far: SO, SO$_2$, DCN, and HDO \citep{Garufi2022,Hsieh2023}, and only one iCOM, formamide \citep{Bianchi2023}.
In this context, it is very important to assess whether the detected methyl formate, dimethyl ether, formic acid, and ketene towards \bhb~can be attributed to shocks caused by gas streamers because this would be the richest gas enrichment observed so far.
On the right side of Fig. \ref{fig: 3comp-ch3och3}, we showed that the emission from hot CH$_3$OCH$_3$ gas presents a 2D Gaussian centroid distribution that is similar to the distribution for hot methanol, indicating that the two species are affected by the same physical process. The moment 0 maps of \chtocht~and \chtocho~presented in Fig. \ref{fig: mom0-ch3och3} and \ref{fig: mom0-ch3ocho} clearly show a shifted peaked emission in the southern region of 11B. This southern emission can be compared with the high spatial resolution dust continuum map from \citet{Alves2019}, where the two hot cores are identified with a complex filamentary structure in their surroundings (right side of Fig. \ref{fig: 3comp-ch3och3}). This supports the idea of the release into the gas phase of molecular species from shocks close to 11B rather than the emission from thermally desorbed species.
In the latter case, the emission should have been symmetric and centred towards the two sources, 11A and 11B, whereas the maps show a compact emission in the southern position alone. On the other hand, the moment 0 map of ketene (see Fig. \ref{fig: spec-h2cco}) shows that the emission is likely peaked in between 11A and 11B. In addition, it shows extended emission that follows the bright southern filament (Fig. \ref{fig: 3comp-ch3och3}) observed by \cite{Alves2019} in the dust continuum. Therefore, ketene could be tracing a more extended region, likely the warm gas streaming in from the circumbinary disk, rather than the shocked regions at the end of the streamer.

The compact emission of iCOMs and the more extended emission of ketene would not be surprising from a chemical point of view.
While iCOMs have very low abundances in cold ($\leq20$ K) gas \citep[e.g.][]{Vastel2014, Jimenez2016}because of their formation routes and binding energies \citep[e.g.][]{Ceccarelli2023}, ketene is observed in cold gas with an abundance about a factor 10 higher than iCOMs \citep[e.g.][]{Jaber2014,Vastel2014}, in agreement with the more extended emission in our observations.

As mentioned in the Introduction, the protobinary system is still deeply embedded in its natal cloud and far from any variable stars. Therefore, although external irradiation could produce shifted emission of molecular species from a protostellar core like this \citep[e.g.][] {Lindberg2012,Lindberg2015}, the location of \bhb~and the derived high gas temperature at which the iCOMs are emitted rules this possibility out.

In summary, our observations suggest that the shocks caused by the streaming gas crashing onto quiescent material surrounding \bhb, likely its circumbinary disk, enrich the gas in iCOMs. We clearly detect methanol, methyl formate, and dimethyl ether, which are among the iCOMs with the highest observed abundances in molecular shocks: X(\chtoh)=(5.0$\pm$1.0)$\times$10$^{-7}$, X(\chtocho)=(2.7$\pm$0.7)$\times$10$^{-8}$, and X(\chtocht)=(2.5$\pm$0.7)$\times$10$^{-8}$ for L1157-B1 \citep{Lefloch2017}, and \chtoh/\chtocht $\sim$ 1 for the IRAS 4A outflow \citep{DeSimone2020-I4outflow}.

\subsection{Comparison with other protostellar sources}

The possible presence of shocks associated with accretion or ejection occurring very close to the central accreting objects and at the origin of iCOM emission, such as in \bhb, has also been invoked towards G328.2551-0.5321 \citet{Csengeri2019}, IRAS16293 A \citep{Maureira2022}, and SVS13A \citep{Bianchi2023}. Towards the high-mass star-forming region G328.2551-0.5321, \citet{Csengeri2019} reported a differential distribution of some O-bearing iCOMs (ethanol, acetone, and ethylene glycol) with respect to some N-bearing iCOMs (vinyl and ethyl cyanide), as found in \bhb, and the other detected iCOMs (e.g. methanol, acetaldehyde, and methyl cyanide). Specifically, the emission from ethanol, acetone, and ethylene glycol peaks roughly perpendicular to the outflow direction and is offset from the continuum emission peak by about 700 au (i.e. one beam), whereas the emission peak of the remaining iCOMs coincides with the dust peak. \citet{Csengeri2019} argued that the iCOM emission offset from the dust peak is due to accretion shocks resulting from the impact of infalling gas. However, higher spatial resolution observations are necessary to resolve the iCOM emission, and thereby, confirm their exact origin (shocks associated with accretion, ejection, or different processes). In addition, our non-detection of C$_2$H$_5$CN and C$_2$H$_3$CN (see the appendix) provides an upper limit to the C$_2$H$_5$CN/CH$_3$OH and C$_2$H$_3$CN/CH$_3$OH abundance ratios of $<$0.07 \% and $<$0.02 \%, respectively, which are much lower than those measured in G328.2551-0.5321, namely 0.57 and 0.17 \%, respectively \citep{Csengeri2019}. Therefore, the comparison between \bhb and G328.2551-0.5321 is inconclusive.\\

In the same context, in IRAS16293 A, \citep{Maureira2022} found hot ($\geq$ [50, 120] K) spots of dust continuum emission, which the authors interpreted as due to shocks from infalling or outflowing gas very close to the central object (on scales of $\sim$10 au). The line emission from formamide and formic acid also seems to be slightly enhanced in these shocks \citep{Maureira2020,Maureira2022}. \\

It is worth mentioning that the dust opacity could be a major actor in the observed apparent spatial distribution of iCOMs, as shown by \citet{Vastel2022,Desimone2020-VLA} (see also the discussion in \cite{Rivilla2017} and \cite{Lee2022}). Specifically, \citet{Desimone2020-VLA} showed that towards the hot corino NGC1333 IRAS 4A1, methanol line emission is easily detected in the radio by VLA, but is undetected in the 1 and 2 millimeter by ALMA and NOEMA \citep[e.g.][]{LopezSepulcre2017}. Both G328.2551-0.5321 and IRAS16293 A have high dust column densities, and the observations were carried out at frequencies of about 350 and 220 GHz, respectively, where the dust opacity is likely to be higher than in NGC1333 IRAS 4A1, especially towards G328.2551-0.5321.

Finally, for SVS13A, \citet{Bianchi2023} imaged (on scales of $\sim$50 au) three features elongated in the direction of known larger-scale streamers of gas falling towards the two central objects of the binary system VLA4A and VLA4B. The features are enriched in formamide (the only iCOM they searched for in their observations), in addition to HDO and SO$_2$. In this case, the image clearly shows formamide emission at the base of the gas streamers. The iCOMs observed towards \bhb~ are likely due to similar shocks.

To evaluate even further whether the iCOMs emission detected towards \bhb~ arise from shocked accreting gas, we list in Table \ref{tab: comp} the abundance ratios with respect to methanol of the iCOMs and some precursors detected in this work, as measured towards \bhb~ and of the solar-like Class 0 and I hot corinos of IRAS 16293 A \citep{Manigand2020} and SVS13A \citep{Bianchi2019, Bianchi2022}, respectively, along with the protostellar shock region L1157-B1 \citep{Lefloch2017}. In the following, we briefly discuss the five species reported in the table.

\begin{table*}
\caption{Comparison of the detected iCOMs and precursors (relative to methanol) in \bhb~with a low-mass protostar \citep[IRAS 16293 A: ][]{Manigand2020}, a Class I protostar \citep[SVS 13-A: ][]{Bianchi2019}, and a shock position towards a low-mass protostar \citep[L1157-B1: ][]{Lefloch2017,Codella2015}.}
\label{tab: comp}
\centering
\small
\begin{tabular}{c c c c c c}
\hline \hline
                         &                        &         &    X(iCOMs)/X(\chtoh)        &     & \\
\hline \hline
Species             &                        &           \bhb          &      IRAS 16293 A    &      L1157-B1  & SVS 13-A\\
Methyl formate    &\chtocho         &   [2.5, 4.5] $\%$   &   $(2\pm 1)\%$        &   (5.4$\pm$1.0)$\%$  & (1.9$\pm$0.4)$\%$\\
Dimethyl ether     & \chtocht         &  [0.5, 0.8] $\%$    &  $(4\pm 2)\%$        &   (4.9$\pm$1.0)$\%$   & (2.1$\pm$0.7)$\%$\\
Formic acid          & t-HCOOH      &   [0.01, 0.5] $\%$  &  $(0.10\pm0.04)\%$         & (1.7$\pm$0.3)$\%$  & $\le 0.07\%$ \\
Ketene                 & H$_2$CCO   &   [0.01, 0.1] $\%$  &  $(0.07\pm 0.03)\%$         &    (1.7$\pm$0.7)$\%$  & (0.19$\pm$0.06)$\%$\\
Acetaldehyde        &  CH$_{3}$CHO    &     $\le$ 1.5$\%$        &   $ (0.03 \pm 0.01)\%$ & (0.6$\pm$0.10)--(4.2$\pm$0.1)$\%$  &  (0.18$\pm$0.11)$\%$         \\
\hline
\end{tabular}
\end{table*}

\paragraph{Methyl formate and dimethyl ether}

The \chtocho/\chtoh~ abundance ratio derived in \bhb~is consistent with that measured in the Class 0 and I hot corinos of Table \ref{tab: comp} (see also Fig. 3 of \cite{Bianchi2019} and Fig. 4 of \cite{vanGelder2020}) as well as towards the shocked region L1157-B1 \citep{Lefloch2017}. Likewise, the \chtocht/\chtoh~ abundance ratio towards \bhb~ is comparable to that measured in other hot corinos \citep[e.g.][]{Bianchi2019, vanGelder2020}.
Conversely, \chtocht/\chtoh~ is slightly lower than that measured towards the shocked regions L1157-B1 \citep[$\sim5\%$:][]{Lefloch2017} and in the outflow shocks of NGC1333 IRAS 4A \citep[$\sim10\%$:][]{DeSimone2020-I4outflow}. The \chtocht/\chtocho~ abundance ratio [0.1, 0.3] in \bhb~lies at the lower end of the correlation observed by several authors between these two molecules \citep{Jaber2014, Brouillet2013, Belloche2020, Coletta2020, Yang2021, Chahine2022}. The observed correlation (\chtocht/\chtocho $\sim$ 1) may imply a sister-sister or mother-daughter relation between methyl formate and dimethyl ether. \cite{Balucani2015} suggested that the second relation applies: Methyl formate is synthesised from methyl ether via a chain of two gas-phase reactions. Alternatively, both species are synthesised on the grain surfaces \citep{Garrod2008}. However, these routes have been studied theoretically and their efficiency is not obvious, given the competition of the combination of radicals with the H atom abstraction in each reaction \citep{Enrique-Romero2022}. However, the recent model by \cite{Garrod2022} needs the introduction of poorly studied diffusive reactions to enhance the dimethyl ether to the observed values. \citet{Chahine2022} measured \chtocht/\chtocho~equal to $\sim0.2$ towards the HOPS-108 hot corino in the OMC-2 FIR 4 protocluster. This value is similar to that measured towards \bhb. 
These authors tried to verify whether a very different cosmic-ray ionisation rate could explain this difference with respect to the other sources. They concluded, however, that cosmic rays do not play a role in shaping the abundance of \chtocht~relative to that of \chtocho~ (see their Fig. B.10). If the two species are linked by gas-phase reactions, another poorly known parameter that could affect the \chtocho/\chtocht~ ratio is the abundance of atomic oxygen,  which may be different in the shocked gas of \bhb~.

\paragraph{Formic acid}
Table \ref{tab: comp} shows that the formic acid versus methanol ratio estimated in our observations is lower by a factor of 3 than the shocked region of L1157 \citep{Lefloch2017}, but comparable to the value found in IRAS 16293 B \citep{Manigand2020}.

\paragraph{Ketene}
Like in the case of formic acid, the H$_2$CCO/\chtoh~ abundance ratio in \bhb~ is more similar to that measured in hot corinos than in the L1157-B1 shock. However, the latter is the only reported ketene detection in a shocked region so far, and it is therefore premature to draw firm conclusions. In addition, we report that in \bhb, ketene emission is more extended than in the other detected iCOMs (see Sec. \ref{sec: results}), which is an argument for an origin in the gas phase likely belonging to the extended envelope surrounding \bhb. As several simple hydrocarbons, ketene is observed in hot and cold gas \citep[e.g.][]{Jaber2014, Zhou2022}. In the gas phase, it is formed from the oxidation of smaller hydrocarbons (e.g. C$_2$H$_3$ + O), while on the grain surface, it can be formed by the reaction of C atoms landing on CO-rich ice \citep{Ferrero2023}. In the case of \bhb, the extended emission would be more easily explained by gas-phase reactions.

\paragraph{Acetaldehyde}
Table \ref{tab: comp} shows that the upper limit that we derived from the observations towards \bhb~ is consistent with the values observed in hot corino IRAS 16293 B \citep{Manigand2020} and the shocked region L1157-B1 \citep{Lefloch2017}. Our $rms$ is too high to be able to explain the non-detection of this species.

\section{Conclusions}

We presented the ALMA observations towards the protobinary system \bhb~within the framework of the FAUST Large Program. We reported the detection of iCOMs and iCOMs precursors, such as \chtocht, \chtocho, t-HCOOH, and H$_2$CCO. The spatial distribution showed that the emission is confined within the ALMA beam, and the velocity distribution shows three different spatial regions. They are comparable with the methanol emission from previous observations of \citet{Vastel2022}. The detected iCOMs are associated with three hot spots, two of which lie close to 11B, and the third hot spot lies near 11A, with a source size of $\sim$ 0.15$^{\prime\prime}$, which is well below the $\sim$ 0.4$^{\prime\prime}$ beam size. These molecular species are likely the result of the shocked material from the incoming filaments that streams towards \bhb~A and B, shown in our integrated intensity maps compared to the dust emission revealing the filamentary structures. 
Radiative transfer modelling was used to reproduce the many detected transitions and compute the abundance ratios with respect to methanol for methyl formate, dimethyl ether, ketene, formic acid, and an upper limit for acetaldehyde.  We then compared the ratios to Class 0 and Class I protostars as well as with the L1157-B1 shocked region. The \chtocho/\chtoh~as well as CH$_{3}$CHO/\chtoh~ratios are similar in all sources, but H$_{2}$CCO/\chtoh~and t-HCOOH/\chtoh~are much lower. Our \chtocht/\chtocho~ratio of [0.1, 0.3] is slightly lower (factor of 3 to 10) than the correlation found in low- and high-mass protostars. 
Our observations suggest that the shocks caused by the streaming gas crashing onto quiescent material surrounding \bhb~can enrich the gas in iCOMs.

\begin{acknowledgements}
This project has received funding within the European Union's Horizon 2020 research and innovation program from the European Research Council (ERC) for the project {\it The Dawn of Organic Chemistry} (DOC, grant agreement No. 741002) and for the project {\it Dust2Planets} (grant agreement No. 101053020), and from the Marie Sklodowska-Curie action for the project {\it Astro-Chemical Origins} (ACO, grant agreement No. 811312). S.B.C. was supported by the NASA Planetary Science Division Internal Scientist Funding Program through the Fundamental Laboratory Research work package (FLaRe). M.B. acknowledges support from the ERC under the European Union's Horizon 2020 research and innovation program MOPPEX 833460. I.J-.S acknowledges funding from grants No. PID2019-105552RB-C41 and PID2022-136814NB-I00 from the Spanish Ministry of Science and Innovation/State Agency of Research MCIN/AEI/10.13039/501100011033 and by {\it ERDF A way of making Europe}. S.Y. thanks the support by Grant-in-Aids from Ministry of Education, Culture, Sports, Science, and Technologies of Japan (18H05222). L.L. acknowledges DGAPA PAPIIT grants IN112820 and IN108324, and CONAHCYT-CF grant 263356. G.S., C.C. and L.P. acknowledge the project PRIN-MUR 2020 MUR BEYOND-2p {\it Astrochemistry beyond the second period elements}, Prot. 2020AFB3FX. G.S. also acknowledges support from the INAF-Minigrant 2023 TRIESTE {\it TRacing the chemIcal hEritage of our originS: from proTostars to planEts}. T.S. thanks the support by Grant-in-Aids from Ministry of Education, Culture, Sports, Science, and Technologies of Japan (20K04025, 20H05845, and 18H05222).
\end{acknowledgements}

\bibliographystyle{aa}
\bibliography{ref}

\clearpage
\onecolumn

\begin{appendix}

\section{Details for the observed transitions}

\begin{longtable}{cccccccc}
\caption{\label{tab: spectro} Spectroscopic parameters for the brightest detected transitions. } \\
\hline\hline
Frequency         &            Quantum numbers                                & $E_{up}$    & $A_{ul}$     &  $g_{up}$ &                    Original beam, $PA$                                                  &   $rms$     &     $\delta V$      \\
  (MHz)              &    ($J, K_{a}, K_{c}, v_{t}$)      &          (K)     &  (s$^{-1}$)   &                 &  ($^{\prime\prime} \times ^{\prime\prime}$,$^{\circ}$)   &     (K)    &  (km s$^{-1}$)  \\
\hline\hline
\endfirsthead
\caption{continued.}\\
\hline\hline
Frequency         &            Quantum numbers                               & $E_{up}$    & $A_{ul}$     &  $g_{up}$ &                    Original beam, $PA$                                                  &   $rms$     &     $\delta V$      \\
  (MHz)              &    ($J, K_{a}, K_{c}, v_{t}$)      &          (K)     &  (s$^{-1}$)   &                 &  ($^{\prime\prime} \times ^{\prime\prime}$,$^{\circ}$)   &     (K)    &  (km s$^{-1}$)  \\
\hline\hline
\endhead
\hline
\endfoot
           &   &                                &  \chtocht\tablefootmark{a}   &   &        &        &   \\
\hline
257915.5190    &  18$_{5,13,0}$ -- 18$_{4,14,0}$      &  190.97   & 8.76~10$^{-5}$  & 222 & 0.44 $\times$ 0.38, -83 & 0.1494 & 0.14 \\
257913.3120    &  18$_{5,13,1}$ -- 18$_{4,14,1}$      &  190.97   & 8.76~10$^{-5}$  & 592 & 0.44 $\times$ 0.38, -83 & 0.1494 & 0.14 \\
257911.1750    &  18$_{5,13,3}$ -- 18$_{4,14,3}$      &  190.97   & 8.76~10$^{-5}$  & 148 & 0.44 $\times$ 0.38, -83 & 0.1494 & 0.14 \\ 
257911.0360   &  18$_{5,13,5}$ -- 18$_{4,14,5}$     &  190.97   & 8.77~10$^{-5}$  & 74 & 0.44 $\times$ 0.38, -83 & 0.1494 &  0.14 \\ 
258549.3080   &  14$_{1,14,0}$ -- 13$_{0,13,0}$     &  93.33   & 1.31~10$^{-4}$  & 290 & 0.44 $\times$ 0.38, -82 & 0.1469 & 0.14 \\ 
258549.0630   &  14$_{1,14,1}$ -- 13$_{0,13,1}$    &  93.33   & 1.31~10$^{-4}$  & 464 & 0.44 $\times$ 0.38, -82 & 0.1469 & 0.14 \\ 
258548.8190   &  14$_{1,14,3}$ -- 13$_{0,13,3}$    &  93.33   & 1.31~10$^{-4}$  & 116 & 0.44 $\times$ 0.38, -82 & 0.1469 & 0.14 \\ 
258548.8190   &  14$_{1,14,5}$ -- 13$_{0,13,5}$    &  93.33   & 1.31~10$^{-4}$  & 174 & 0.44 $\times$ 0.38, -82 & 0.1469 & 0.14 \\ 
\hline
                           &                                     &                                &  \chtocho\tablefootmark{b}   &   &        &     &      \\
\hline
233212.773        & 19$_{4,16,1}$ -- 18$_{4,15,1}$  & 123.26 & 1.82~10$^{-4}$  & 78 & 0.38 $\times$ 0.34, 70 & 0.115 & 0.63 \\
233226.788        & 19$_{4,16,0}$ -- 18$_{4,15,0}$  & 123.25 & 1.82~10$^{-4}$  & 78 & 0.38 $\times$ 0.34, 70 & 0.115 & 0.63 \\
233649.88          & 19$_{12,7,2}$ -- 18$_{12,6,2}$  & 207.62 & 1.16~10$^{-4}$  & 78 & 0.38 $\times$ 0.34, 70 & 0.115 & 0.63 \\
233655.342        & 19$_{12,8,0}$ -- 18$_{12,7,0}$ & 207.62 & 1.16~10$^{-4}$  & 78 & 0.38 $\times$ 0.34, 70 & 0.115 & 0.63 \\
233655.342        & 19$_{12,7,0}$ -- 18$_{12,6,0}$ & 207.62 & 1.16~10$^{-4}$  & 78 & 0.38 $\times$ 0.34, 70 & 0.115 & 0.63 \\
233670.980        & 19$_{12,8,1}$ --  18$_{12,7,1}$ & 207.60  & 1.16~10$^{-4}$  & 78 & 0.38 $\times$ 0.34, 70 & 0.115 & 0.63 \\
233753.960         & 18$_{4,14,2}$ -- 17$_{4,13,2}$ & 114.37  & 1.84~10$^{-4}$  & 74 & 0.38 $\times$ 0.34, 70 & 0.115 & 0.63 \\
233777.521         & 18$_{4,14,0}$ -- 17$_{4,13,0}$ & 114.36  & 1.84~10$^{-4}$  & 74 & 0.38 $\times$ 0.34, 70 & 0.115 & 0.63 \\
233845.233         & 19$_{11,8,2}$ -- 18$_{11,7,2}$ & 192.39  & 1.29~10$^{-4}$  & 78 & 0.38 $\times$ 0.34, 70 & 0.115 & 0.63 \\
233854.286         & 19$_{11,9,0}$ -- 18$_{11,8,0}$ & 192.39  & 1.29~10$^{-4}$  & 78 & 0.38 $\times$ 0.34, 70 & 0.115 & 0.63 \\
233854.286         & 19$_{11,8,0}$ -- 18$_{11,7,0}$ & 192.39  & 1.29~10$^{-4}$  & 78 & 0.38 $\times$ 0.34, 70 & 0.115 & 0.63 \\
233867.193         & 19$_{11,9,1}$ -- 18$_{11,8,1}$ & 192.38  & 1.29~10$^{-4}$  & 78 & 0.38 $\times$ 0.34, 70 & 0.115 & 0.63 \\
234112.330         & 19$_{10,9,2}$ -- 18$_{10,8,2}$ & 178.51  & 1.40~10$^{-4}$  & 78 & 0.38 $\times$ 0.34, 70 & 0.115 & 0.63 \\
234124.883         & 19$_{10,9,0}$ -- 18$_{10,8,0}$ & 178.50  & 1.40~10$^{-4}$  & 78 & 0.38 $\times$ 0.34, 70 & 0.115 & 0.63 \\
234124.883         & 19$_{10,10,0}$ -- 18$_{10,9,0}$ & 178.50  & 1.40~10$^{-4}$  & 78 & 0.38 $\times$ 0.34, 70 & 0.115 & 0.63 \\
234134.600         & 19$_{10,10,1}$ -- 18$_{10,9,1}$ & 178.50  & 1.40~10$^{-4}$  & 78 & 0.38 $\times$ 0.34, 70 & 0.115 & 0.63 \\
234486.395         & 19$_{9,10,2}$ -- 18$_{9,9,2}$ & 165.98  & 1.51~10$^{-4}$  & 78 & 0.38 $\times$ 0.34, 70 & 0.115 & 0.63 \\
234502.2407        & 19$_{9,11,0}$ -- 18$_{9,10,0}$ & 165.97  & 1.51~10$^{-4}$  & 78 & 0.38 $\times$ 0.34, 70 & 0.115 & 0.63 \\
234502.4315        & 19$_{9,10,0}$ -- 18$_{9,9,0}$ & 165.97  & 1.51~10$^{-4}$  & 78 & 0.38 $\times$ 0.34, 70 & 0.115 & 0.63 \\
234508.614        & 19$_{9,11,1}$ -- 18$_{9,10,1}$ & 165.97  & 1.51~10$^{-4}$  & 78 & 0.38 $\times$ 0.34, 70 & 0.115 & 0.63 \\
245883.179        & 20$_{13,7,2}$ -- 19$_{13,6,2}$ & 235.98  & 1.30~10$^{-4}$  & 82 & 0.46 $\times$ 0.39, -80 & 0.0373 & 1.19 \\
245885.243       & 20$_{13,7,0}$ -- 19$_{13,6,0}$ & 235.98  & 1.30~10$^{-4}$  & 82 & 0.46 $\times$ 0.39, -80 & 0.0373 & 1.19 \\
245885.243       & 20$_{13,8,0}$ -- 19$_{13,7,0}$ & 235.98  & 1.30~10$^{-4}$  & 82 & 0.46 $\times$ 0.39, -80 & 0.0373 & 1.19 \\
245903.680      & 20$_{13,8,1}$ -- 19$_{13,7,1}$ & 235.97  & 1.30~10$^{-4}$  & 82 & 0.46 $\times$ 0.39, -80 & 0.0373 & 1.19 \\
246187.016     & 21$_{2,19,3}$ -- 20$_{2 ,18,3}$ & 326.63  & 2.18~10$^{-4}$  & 86 & 0.46 $\times$ 0.39, -80 & 0.0373 & 1.19 \\
246600.012     & 20$_{10,10,2}$ -- 19$_{10,9,2}$ & 190.35  & 1.70~10$^{-4}$  & 82 & 0.46 $\times$ 0.39, -80 & 0.0373 & 1.19 \\
246613.392    & 20$_{10,10,0}$ -- 19$_{10,9,0}$ & 190.34  & 1.70~10$^{-4}$  & 82 & 0.46 $\times$ 0.39, -80 & 0.0373 & 1.19 \\
246613.392    & 20$_{10,11,0}$ -- 19$_{10,10,0}$ & 190.34  & 1.70~10$^{-4}$  & 82 & 0.46 $\times$ 0.39, -80 & 0.0373 & 1.19 \\
246615.129    & 20$_{5,16,3}$ -- 19$_{5,15,3}$ & 328.24  & 2.12~10$^{-4}$  & 82 & 0.46 $\times$ 0.39, -80 & 0.0373 & 1.19 \\
246623.190   & 20$_{10,11,1}$ -- 19$_{10,10,1}$ & 190.33  & 1.70~10$^{-4}$  & 82 & 0.46 $\times$ 0.39, -80 & 0.0373 & 1.19 \\
246891.611   & 19$_{4,15,2}$ -- 18$_{4,14,2}$ & 126.22  & 2.18~10$^{-4}$  & 78 & 0.46 $\times$ 0.39, -80 & 0.0373 & 1.19 \\
246914.658   & 19$_{4,15,0}$ -- 18$_{4,14,0}$ & 126.21  & 2.18~10$^{-4}$  & 78 & 0.46 $\times$ 0.39, -80 & 0.0373 & 1.19 \\
247040.65   & 20$_{9,11,2}$ -- 19$_{9,10,2}$ & 177.84  & 1.82~10$^{-4}$  & 82 & 0.46 $\times$ 0.39, -80 & 0.0373 & 1.19 \\
247044.146   & 21$_{3,19,1}$ -- 20$_{3,18,1}$ & 139.90  & 2.21~10$^{-4}$  & 86 & 0.46 $\times$ 0.39, -80 & 0.0373 & 1.19 \\
247053.453   & 21$_{3,19,0}$ -- 20$_{3,18,0}$ & 139.89  & 2.21~10$^{-4}$  & 86 & 0.46 $\times$ 0.39, -80 & 0.0373 & 1.19 \\
247057.2591   & 20$_{9,12,0}$ -- 19$_{9,11,0}$ & 177.83  & 1.82~10$^{-4}$  & 82 & 0.46 $\times$ 0.39, -80 & 0.0373 & 1.19 \\
247057.7373   & 20$_{9,11,0}$ -- 19$_{9,10,0}$ & 177.83  & 1.82~10$^{-4}$  & 82 & 0.46 $\times$ 0.39, -80 & 0.0373 & 1.19 \\
247063.662   & 20$_{9,12,1}$ -- 19$_{9,11,1}$   & 177.82  & 1.82~10$^{-4}$  & 82 & 0.46 $\times$ 0.39, -80 & 0.0373 & 1.19 \\
247469.2492   & 23$_{1,23,3}$ -- 22$_{1,22,3}$   & 332.88  & 2.28~10$^{-4}$  & 94 & 0.46 $\times$ 0.39, -80 & 0.0373 & 1.19 \\
247469.5853   & 23$_{0,23,3}$ -- 22$_{0,22,3}$   & 332.88  & 2.28~10$^{-4}$  & 94 & 0.46 $\times$ 0.39, -80 & 0.0373 & 1.19 \\
247516.3123   & 23$_{0,23,5}$ -- 22$_{0,22,5}$   & 332.13  & 2.29~10$^{-4}$  & 94 & 0.46 $\times$ 0.39, -80 & 0.0373 & 1.19 \\
247516.0121   & 23$_{1,23,4}$ -- 22$_{1,22,4}$   & 332.13  & 2.29~10$^{-4}$  & 94 & 0.46 $\times$ 0.39, -80 & 0.0373 & 1.19 \\
257919.89   & 21$_{16,5,2}$ -- 20$_{16,4,2}$   & 306.01  & 1.09~10$^{-4}$  & 86 & 0.44 $\times$ 0.38, -83 & 0.1494 & 0.14 \\
257910.566  & 21$_{16,5,0}$ -- 20$_{16,4,0}$   & 306.01  & 1.09~10$^{-4}$  & 86 & 0.44 $\times$ 0.38, -83 & 0.1494 & 0.14 \\
257910.566  & 21$_{16,6,0}$ -- 20$_{16,5,0}$   & 306.01  & 1.09~10$^{-4}$  & 86 & 0.44 $\times$ 0.38, -83 & 0.1494 & 0.14 \\
257906.128  & 21$_{8,13,3}$ -- 20$_{8,12,3}$   & 365.98  & 2.22~10$^{-4}$  & 86 & 0.44 $\times$ 0.38, -83 & 0.1494 & 0.14 \\
257889.872  & 21$_{8,14,3}$ -- 20$_{8,13,3 }$  & 365.98  & 2.22~10$^{-4}$  & 86 & 0.44 $\times$ 0.38, -83 & 0.1494 & 0.14 \\
\hline
                           &                                     &                                &  H$_2$CCO\tablefootmark{c}   &   &        &     &      \\
\hline
260191.982  & 13$_{1,13}$ -- 12$_{1,12}$  & 100.5  & 1.98~10$^{-4}$  & 81 & 0.44 $\times$ 0.37, -83 & 0.1071 & 0.14 \\
\hline
                           &                                     &                                &  t-HCOOH\tablefootmark{d}   &   &        &     &      \\
\hline
246105.9739  & 11$_{2,10}$--10$_{2,9}$  & 83.74   & 1.62~10$^{-4}$  & 23 & 0.46 $\times$ 0.39, -80 & 0.0373 & 1.19 \\
247446.2439  & 11$_{6,6}$--10$_{6,5}$    & 185.63 & 1.20~10$^{-4}$  & 23 & 0.46 $\times$ 0.39, -80 & 0.0373 & 1.19 \\
247513.9713  & 11$_{5,7}$--10$_{5,6}$    & 150.71 & 1.35~10$^{-4}$   & 23 & 0.46 $\times$ 0.39, -80 & 0.0373 & 1.19 \\
247514.1176  & 11$_{5,6}$--10$_{5,5}$    & 150.71 & 1.35~10$^{-4}$   & 23 & 0.46 $\times$ 0.39, -80 & 0.0373 & 1.19 \\
\end{longtable} 
\tablefoot{
\tablefoottext{a}{The spectroscopic parameters for \chtocht~come from CDMS (TAG=46514) \citep{Endres2009}. The AA, EE, EA, and AE substates are represented by the $v_{t}$ designations of 0, 1, 3, and 5, respectively.}\\
\tablefoottext{b}{The spectroscopic parameters for \chtocho~come from JPL (TAG=60003)] \citep{Ilyushin2009} with A$_\mathrm{ul}$ larger than 10$^{-4}$ s$^{-1}$.}\\
\tablefoottext{c}{The spectroscopic parameters for H$_{2}$CCO come from CDMS (TAG=42501) \citep{Brown1990}.}\\
\tablefoottext{d}{The spectroscopic parameters for  t-HCOOH come from CDMS (TAG=46506) \citep{Winnewisser2002}.}\\
} 
\twocolumn

\section{Upper limits of $\rm N(C_2H_5CN)$ and $\rm N(C_2H_3CN)$}

\begin{figure}
    \centering
    \includegraphics[width=0.55\textwidth]{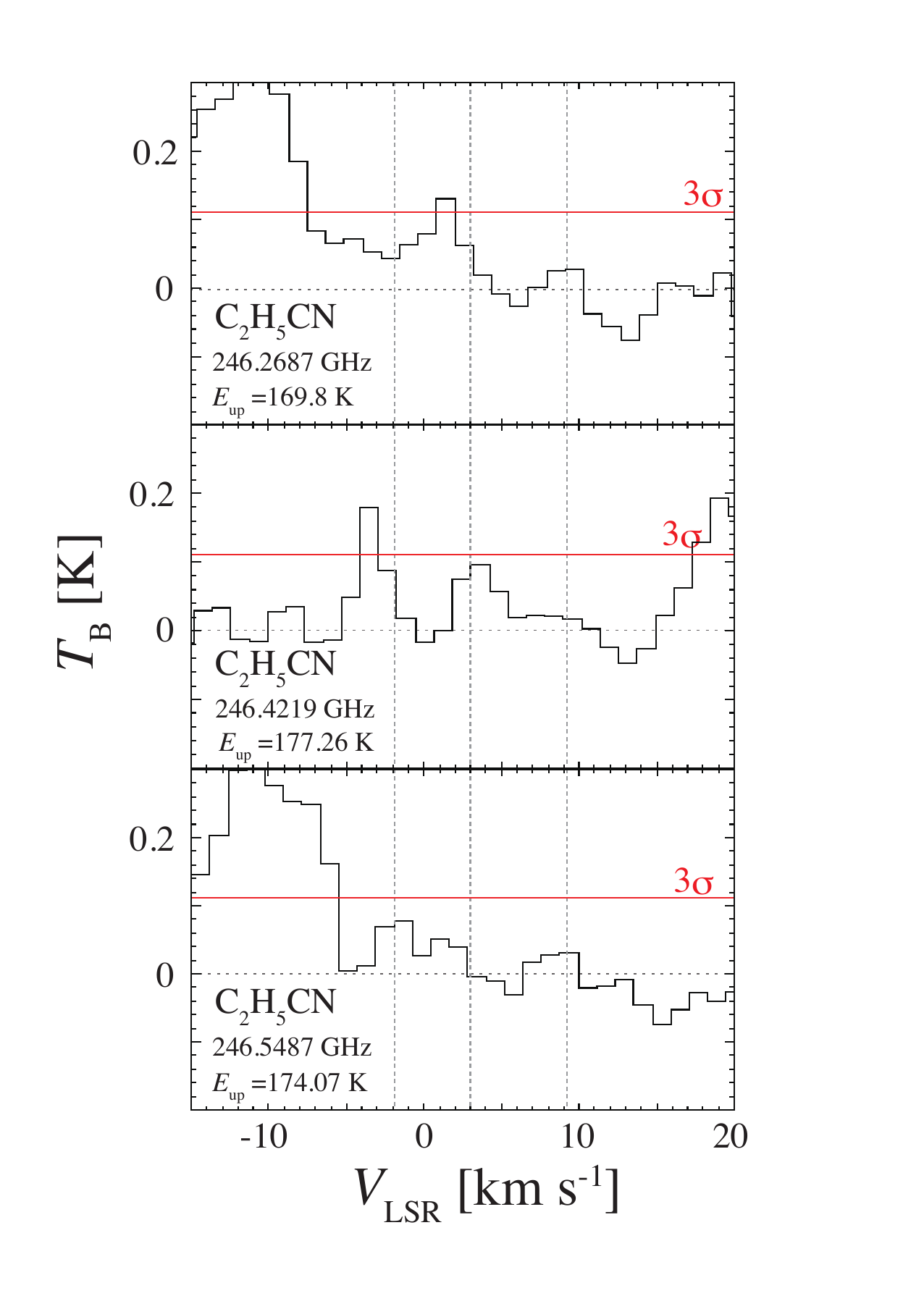}
    \caption{C$_2$H$_5$CN transitions. The vertical dotted lines indicate the velocity of the three velocity components (-2, 2.8, and 9.9 km s$^{-1}$).  The red lines indicate the 3 $\sigma$ noise level. \label{fig: a1}}
\end{figure}

\begin{figure}
    \centering
    \includegraphics[width=0.55\textwidth]{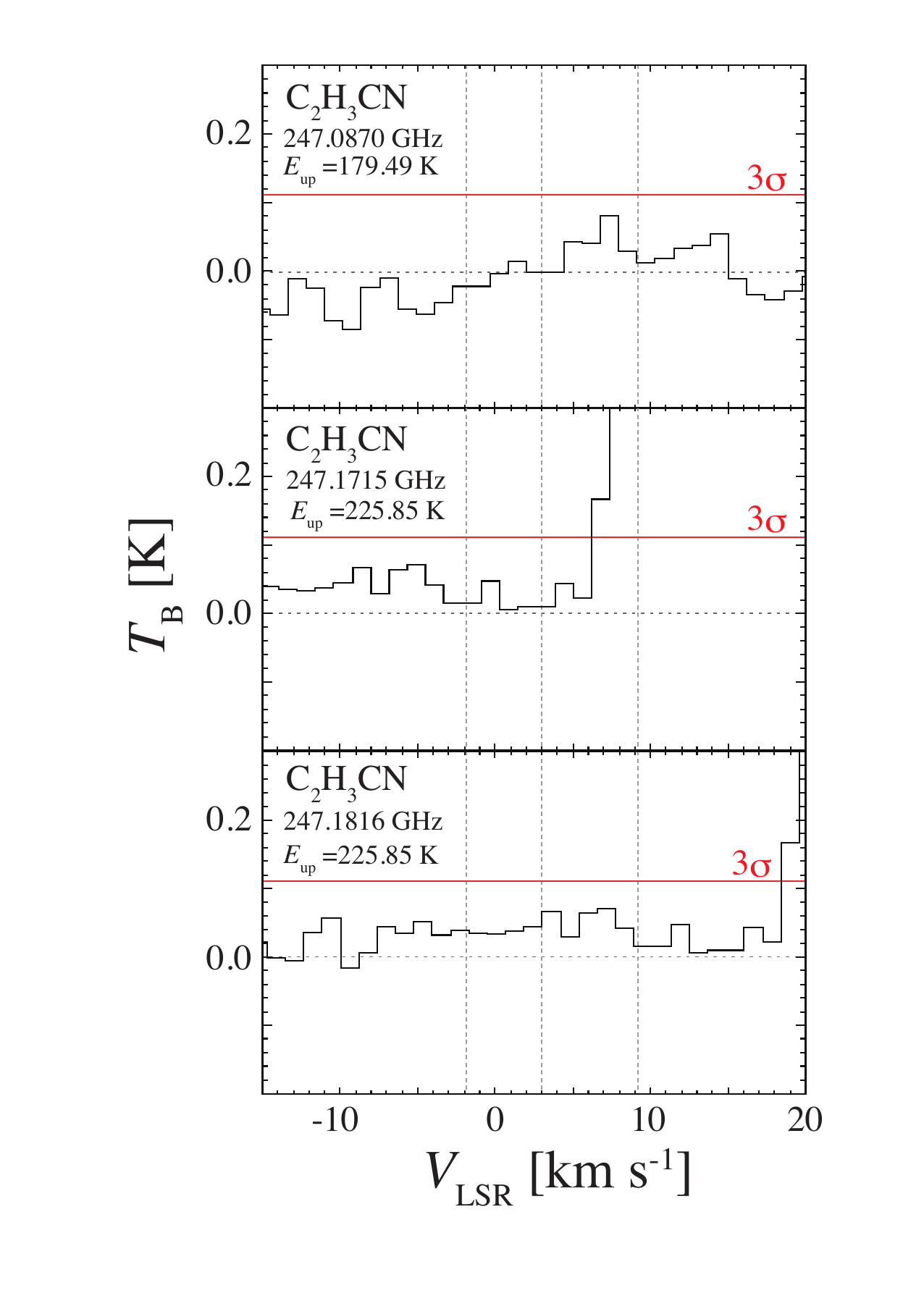}
\caption{C$_2$H$_3$CN transitions. The vertical dotted lines indicate the velocity of the three velocity components (-2, 2.8, and 9.9 km s$^{-1}$). The red lines indicate the 3 $\sigma$ noise level. \label{fig: a2}}
\end{figure}

We present in this section the computation for the upper limits on the nitrogen bearing iCOMs. \\
Figure \ref{fig: a1} shows the observed spectra of three C$_2$H$_5$CN lines within 0.82$^{\prime\prime}$$\times$0.77$^{\prime\prime}$ centred at $\alpha$(2000) = 17h11m23.107s, $\delta$(2000) = -27$^{\circ}$24$^{\prime}$33.005$^{\prime\prime}$. The quantum numbers (and spectroscopic parameters from CDMS) associated for the three transitions are listed in Table \ref{tab: spectroND}. Of the three transitions, the 28$_{2,27}$--26$_{2,26}$ line is expected to be strongest. Some features can be seen in Figure \ref{fig: a1}, but they are all offset from the velocity of the three velocity components (-2, 2.8, and 9.9 km~s$^{-1}$). We can safely conclude that the C$_2$H$_5$CN emission is not detected within the noise. \\
Figure \ref{fig: a2} shows the observed spectra of three C$_2$H$_3$CN lines whose quantum numbers (and spectroscopic parameters from CDMS) are listed in Table \ref{tab: spectroND}. Of the three transitions, the C$_2$H$_3$CN 26$_{3,24}$--25$_{3,23}$ line is expected to be strongest. In Figure \ref{fig: a2}, we can safely conclude that no C$_2$H$_3$CN emission is detected within the noise. From the observed transitions of C$_2$H$_5$CN 28$_{2,27}$--26$_{2,26}$ and C$_2$H$_3$CN 26$_{3,24}$--25$_{3,23}$, we can evaluate an upper limit on their column densities. For the derivation, we assumed an LTE condition and used the following equation:
\begin{equation}
N_{total}=\frac{3h}{8\pi^3 \mu^2 S f}Q \frac{\exp\left(\frac{E_{u}}{kT_{\rm ex}}\right)}{\exp\left(\frac{h \nu}{kT_{\rm ex}}\right)-1} \frac{W_u}{J(T_{\rm ex})-J(T_{\rm BB})},
\end{equation}
where $h$ is the Planck constant, $Q$ is the partition function, $\nu$ is the rest frequency, $\mu$ is the electric dipole moment, $S$ is the line strength, $f$ is the beam filling factor, $E_{up}$ is the upper state energy, $T_{\rm ex}$ is the excitation temperature, $T_{\rm BB}$ is the temperature of the cosmic background radiation (2.73 K), $W_u$ is the upper limit of integrated intensity, $k$ is the Boltzmann constant, and $J$ is the Planck function for the given temperature and frequency ($\nu$) as
\begin{equation}
J(T) = \frac{\frac{h\nu}{k}}{\exp\left(\frac{h\nu}{kT}\right)-1}.
\end{equation}
The upper limit of the integrated intensity (in K~km~s$^{-1}$) was calculated from the following equation:
\begin{equation}
W_{u}= \sigma \delta V \sqrt{\frac{\Delta V}{\delta V}},
\end{equation}
where $\sigma$ is the $rms$ noise level of the spectra at the channel width of $\delta V$ (Table \ref{tab: spectro}), and $\Delta V$ is the velocity width of a target. For the calculations, we adopted the parameters from Table \ref{tab: a1}. 
The partition function $Q$ was evaluated using the values reported in the CDMS database. We fit the CDMS data to the function of $Q$($T$) = $a$$\times$$T^b$, where $a$ and $b$ are fitting parameters. For C$_2$H$_5$CN, ($a$, $b$)=(7.2158, 1.4995). For C$_2$H$_3$CN, ($a$, $b$)=(0.95923, 2.0865). The derived upper limits are presented in Table \ref{tab: a1}.

\begin{table}
\small
\caption{Upper limits for $N$(C$_2$H$_5$CN) and $N$(C$_2$H$_3$CN).}
\label{tab: a1}
\begin{tabular}{c c c c}
\hline
  &  & Velocity components &\\
\hline 
             &  -2 km~s$^{-1}$ & 2.8 km~s$^{-1}$  & 9.9 km~s$^{-1}$   \\
\hline 
$N$(C$_2$H$_5$CN) (cm$^{-2}$) & $<$3.5$\times$10$^{14}$ &  $<$1.9$\times$10$^{15}$  & $<$2.0$\times$10$^{15}$ \\
$N$(C$_2$H$_3$CN) (cm$^{-2}$) & $<$2.4$\times$10$^{14}$ &  $<$5.2$\times$10$^{14}$  & $<$5.4$\times$10$^{14}$ \\
$\theta_S$ ($^{\prime\prime}$) & 0.15 & 0.13 & 0.12 \\
$\Delta V$ (km s$^{-1}$) & 2 & 5 & 4 \\
Temperature (K) & 80 & 80 & 80\\ 
\hline
\end{tabular}
\tablefoot{The identification of the three velocity components is based on the methanol analysis described in \citet{Vastel2022}.}
\end{table}

\newpage

\section{Spectroscopic parameters for the non-detected transitions}

\begin{table*}
\small
\centering
\caption{Spectroscopic parameters for some non-detected transitions we used to compute upper limits.}
\label{tab: spectroND}
\begin{tabular}{ccccccccc}
\hline
Frequency         &            Quantum numbers                    & $E_{up}$    & $A_{ul}$     &  $g_{up}$ & Original beam, $PA$ &   $rms$\tablefootmark{a}     &     $\delta V$      \\
  (MHz)              &    ($J, K_{a}, K_{c}, v_{t}$)      &          (K)                  &       (s$^{-1}$)         & &  ($^{\prime\prime} \times ^{\prime\prime}$,$^{\circ}$)   &     (K)    &  (km s$^{-1}$)  \\
\hline
                           &                                     &                                &  CH$_{3}$CHO\tablefootmark{b}    &   &        &     &      \\
\hline
232980.5688  & 5$_{2,3,2}$ -- 4$_{1,3,2}$  & 23.03  & 1.34~10$^{-5}$  & 22 & 0.38 $\times$ 0.34, 70 & 0.115 & 0.63 \\
234469.2610  & 3$_{3,1,1}$ -- 3$_{2,2,1}$  & 25.84  & 2.08~10$^{-5}$  & 14 & 0.38 $\times$ 0.34, 70 & 0.115 & 0.63 \\
234241.5647  & 3$_{3,0,2}$ -- 3$_{2,1,2}$  & 25.94  & 2.08~10$^{-5}$  & 14 & 0.38 $\times$ 0.34, 70 & 0.115 & 0.63 \\
247432.9970  & 6$_{2,5,1}$ -- 5$_{1,4,2}$  & 28.46  & 2.48~10$^{-5}$  & 26 & 0.46 $\times$ 0.39, -80 & 0.0373 & 1.19 \\
234437.9503  & 4$_{3,2,1}$ -- 4$_{2,3,1}$  & 29.53  & 2.92~10$^{-5}$  & 18 & 0.38 $\times$ 0.34, 70 & 0.115 & 0.63 \\
234198.3312  & 4$_{3,1,2}$ -- 4$_{2,2,2}$  & 29.64  & 2.91~10$^{-5}$  & 18 & 0.38 $\times$ 0.34, 70 & 0.115 & 0.63 \\
234396.7456  & 5$_{3,3,1}$ -- 5$_{2,4,1}$  & 34.16  & 3.30~10$^{-5}$  & 22 & 0.38 $\times$ 0.34, 70 & 0.115 & 0.63 \\
234091.1815  & 5$_{3,2,2}$ -- 5$_{2,3,2}$  & 34.26  & 3.29~10$^{-5}$  & 22 & 0.38 $\times$ 0.34, 70 & 0.115 & 0.63 \\
234385.7553  & 6$_{3,4,1}$ -- 6$_{2,5,1}$  & 39.71  & 3.40~10$^{-5}$  & 26 & 0.38 $\times$ 0.34, 70 & 0.115 & 0.63 \\
233845.3823  & 6$_{3,3,2}$ -- 6$_{2,4,2}$  & 39.81  & 3.38~10$^{-5}$  & 26 & 0.38 $\times$ 0.34, 70 & 0.115 & 0.63 \\
234486.2127  & 7$_{3,5,1}$ -- 7$_{2,6,1}$  & 46.19  & 3.26~10$^{-5}$  & 30 & 0.38 $\times$ 0.34, 70 & 0.115 & 0.63 \\
234144.5136  & 7$_{3,4,0}$ -- 7$_{2,5,0}$  & 46.29  & 3.80~10$^{-5}$  & 30 & 0.38 $\times$ 0.34, 70 & 0.115 & 0.63 \\
233341.7003  & 7$_{3,4,2}$ -- 7$_{2,5,2}$  & 46.29  & 3.22~10$^{-5}$  & 30 & 0.38 $\times$ 0.34, 70 & 0.115 & 0.63 \\
233124.9632  & 8$_{3,5,0}$ -- 8$_{2,6,0}$  & 53.69  & 3.88~10$^{-5}$  & 34 & 0.38 $\times$ 0.34, 70 & 0.115 & 0.63 \\
247142.155  & 14$_{0,14,0}$ -- 13$_{1,13,0}$  & 95.68  & 6.64~10$^{-5}$  & 58 & 0.46 $\times$ 0.39, -80 & 0.0373 & 1.19 \\
247341.3322  & 14$_{0,14,2}$ -- 13$_{1,13,1}$  & 95.76  & 6.63~10$^{-5}$  & 58 & 0.46 $\times$ 0.39, -80 & 0.0373 & 1.19 \\
246330.7304  & 15$_{3,13,0}$ -- 15$_{2,14,0}$  & 131.49  & 4.89~10$^{-5}$  & 62 & 0.46 $\times$ 0.39, -80 & 0.0373 & 1.19 \\
234401.4565  & 18$_{2,17,1}$ -- 18$_{1,18,1}$  & 166.47  & 2.71~10$^{-5}$  & 74 & 0.38 $\times$ 0.34, 70 & 0.115 & 0.63 \\
245939.644  & 19$_{2,18,0}$ -- 19$_{1,19,0}$  & 183.86  & 3.04~10$^{-5}$  & 78 & 0.46 $\times$ 0.39, -80 & 0.0373 & 1.19 \\
233048.5159  & 12$_{2,11,4}$ -- 11$_{2,10,4}$  & 285.13  & 4.15~10$^{-5}$  & 50 & 0.38 $\times$ 0.34, 70 & 0.115 & 0.63 \\
\hline
                           &                                     &                                &  NH$_2$CHO\tablefootmark{c}    &   &        &     &      \\
\hline
247390.719  & 12$_{0,12}$ -- 11$_{0,11}$  & 78.12  & 1.10~10$^{-3}$  & 25 & 0.46 $\times$ 0.39, -80 & 0.0373 & 1.19 \\
260189.0898  & 12$_{2,10}$ -- 11$_{2,9}$  & 92.35  & 1.25~10$^{-3}$  & 25 & 0.44 $\times$ 0.37, -83 & 0.1071 & 0.14 \\
233896.577  & 11$_{3,9}$ -- 10$_{3,8}$  & 94.11  & 8.62~10$^{-4}$  & 23 & 0.38 $\times$ 0.34, 70 & 0.115 & 0.63 \\
234315.498  & 11$_{3,8}$ -- 10$_{3,7}$  & 94.16  & 8.67~10$^{-4}$  & 23 & 0.38 $\times$ 0.34, 70 & 0.115 & 0.63 \\
\hline
                           &                                     &                                &  CH$_{3}$O$^{13}$CHO\tablefootmark{d}    &   &        &     &      \\
\hline
246787.916  & 23$_{1,23,1}$ -- 22$_{1,22,1}$  & 144.64  & 2.31~10$^{-4}$  & 94 & 0.46 $\times$ 0.39, -80 & 0.0373 & 1.19 \\
246788.142  & 23$_{0,23,1}$ -- 22$_{0,22,1}$  & 144.64  & 2.31~10$^{-4}$  & 94 & 0.46 $\times$ 0.39, -80 & 0.0373 & 1.19 \\
246788.838  & 23$_{1,23,0}$ -- 22$_{1,22,0}$  & 144.62  & 2.31~10$^{-4}$  & 94 & 0.46 $\times$ 0.39, -80 & 0.0373 & 1.19 \\
246789.063  & 23$_{0,23,0}$ -- 22$_{0,22,0}$  & 144.62  & 2.31~10$^{-4}$  & 94 & 0.46 $\times$ 0.39, -80 & 0.0373 & 1.19 \\
\hline

                           &                                     &                                &  C$_2$H$_5$CN   &   &        &     &      \\
\hline
246268.7369      &   27$_{2,25}$--26$_{2,24}$  &  169.80  & 1.26 10$^{-3}$  & 55  & 0.46 $\times$ 0.39, -80 & 0.0373 & 1.19 \\
246421.9174      &   28$_{2,27}$--26$_{2,26}$  &  177.26  & 1.26 10$^{-3}$  & 57  &  0.46 $\times$ 0.39, -80 & 0.0373 & 1.19 \\
246548.6987      &   27$_{3,24}$-26$_{3,23}$   &  174.065 & 1.25 10$^{-3}$  & 55 & 0.46 $\times$ 0.39, -80 & 0.0373 & 1.19 \\
\hline
                           &                                     &                                &  C$_2$H$_3$CN   &   &        &     &      \\
\hline
247087.0333      &    26$_{3,24}$--25$_{3,23}$  &  179.49  &  3.33 10$^{-3}$ & 159  & 0.46 $\times$ 0.39, -80 & 0.0373 & 1.19 \\
247171.5135      &    25$_{6,20}$--26$_{5,21}$  &   225.85 &  3.17 10$^{-5}$  & 153  & 0.46 $\times$ 0.39, -80 & 0.0373 & 1.19 \\
247181.6118      &    25$_{6,19}$--26$_{5,22}$  &   225.85 &  3.17 10$^{-5}$    & 153  & 0.46 $\times$ 0.39, -80 & 0.0373 & 1.19 \\
\hline              
\end{tabular}
\tablefoot{
\tablefoottext{a}{The $rms$ is computed from the spectra extracted within $0.82^{\prime\prime} \times 0.77^{\prime\prime}$.}\\
\tablefoottext{b}{The spectroscopic parameters for CH$_{3}$CHO \citep{Kleiner1996} comes from JPL (TAG=44003) with A$_\mathrm{ul}$ larger than 10$^{-5}$ s$^{-1}$. }\\
\tablefoottext{c}{The spectroscopic parameters for NH$_2$CHO \citep{Kryvda2009} comes from CDMS (TAG=45512) with A$_\mathrm{ul}$ larger than 10$^{-4}$ s$^{-1}$. }\\
\tablefoottext{d}{The spectroscopic parameters for CH$_{3}$O$^{13}$CHO \citep{Carvajal2010} comes from CDMS (TAG=61515) with A$_\mathrm{ul}$ larger than 10$^{-4}$ s$^{-1}$.  }\\
}
\end{table*}

\begin{figure}
    \centering
    \includegraphics[width=0.45\textwidth]{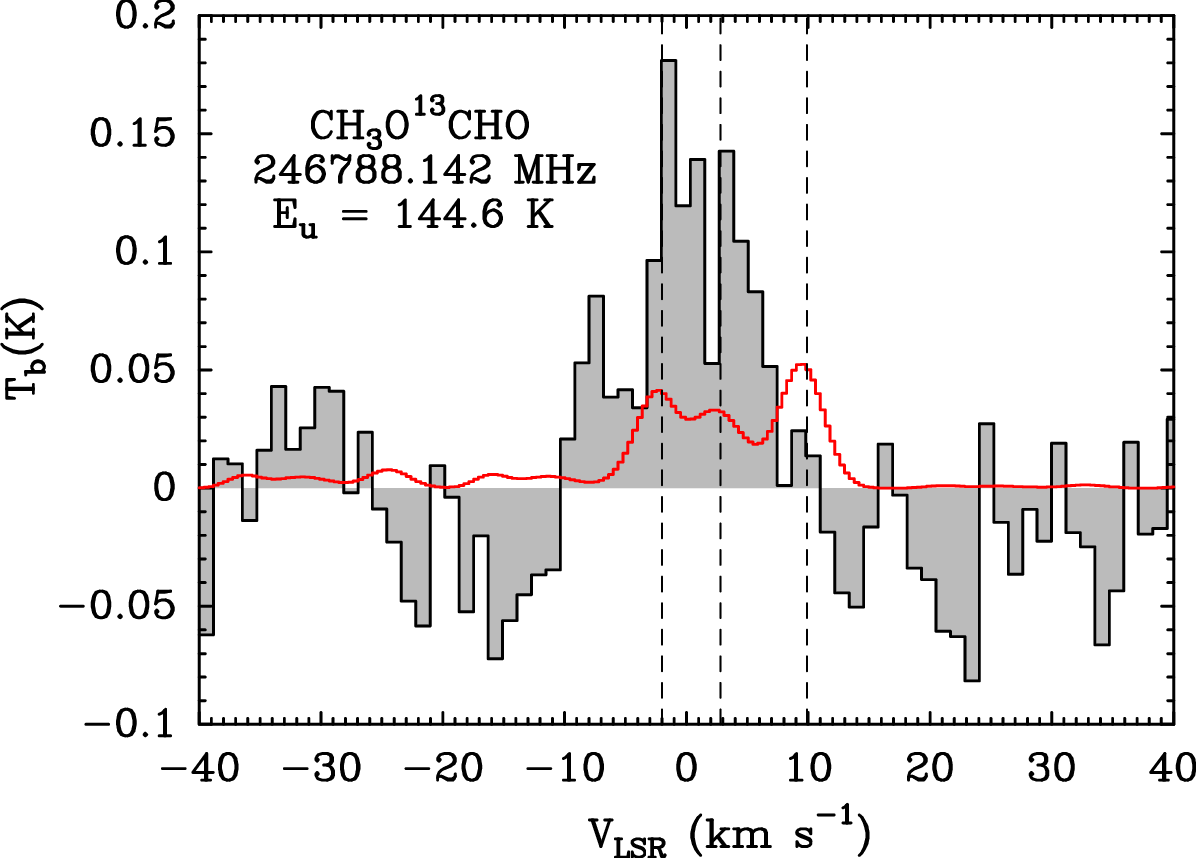}
    \caption{Potential CH$_{3}$O$^{13}$CHO transitions at 246.78 GHz whose spectroscopic parameters are listed in Table \ref{tab: spectro}. The dashed lines correspond to the results of the best-fit Gaussian components obtained from the high spectral resolution methanol transitions at -2 km~s$^{-1}$, 2.8 km~s$^{-1}$ and 9.9 km~s$^{-1}$ \citep{Vastel2022}. The red line corresponds to the modelled transitions as described in Sect. \ref{sec: modelling}. The shaded area corresponds to the filled histogram above and below the baseline.}
    \label{fig: ch3o13cho}
\end{figure}

\end{appendix}

\end{document}